\documentclass[english]{revtex4-1}
\usepackage[T1]{fontenc}
\usepackage[latin9]{inputenc}
\usepackage{array}
\usepackage{booktabs}
\usepackage{calc}
\usepackage{textcomp}
\usepackage{multirow}
\usepackage{amsmath}
\usepackage{graphicx}
\usepackage{esint}

\makeatletter

\providecommand{\tabularnewline}{\\}

\makeatother

\usepackage{babel}
\begin{document}

\title{Thermoelectric Thomson's relations revisited for a linear energy
converter}

\author{S. Gonzalez\textendash Hernandez and L. A. Arias\textendash Hernandez}

\address{Dpto. de Física, Escuela Superior de Física y Matemáticas, Instituto
Politécnico Nacional, U. P. Adolfo López Mateos, Zacatenco, C.P. 07738,
México D.F., México, Email: samayoa126@hotmail.com}

\date{19/dic/2017}
\begin{abstract}
In this paper we revisit the thermocouple model, as a linear irreversible
thermodynamic energy converter. As is well known, the linear model
of the thermocuple is one of the classics in this branch. In this
model we note two types of phenomenological coefficients: the first
comes from some microscopic models, such as the coefficient associated
with the electric conductivity, and the second comes from experimental
facts such as the coefficient associated with the thermoelectric power.
We show that in the last case, these coefficients can be related to
the operation modes of the converter. These relationships allow us
to propose a generalization of the first and second Thomson's relations.
For this purpose we develop the ideas of non-isothermal linear converters,
operated directly (heat engine) and indirect (refrigerator). In addition
to this development we analyze the energy described by these converters.
\end{abstract}
\maketitle

\section{Introduction}

Thermoelectricity is a seminal phenomenon in Non-Equilibrium Thermodynamics;
within the effects that constitute this phenomenon, three are well
known T. J. Seebeck discovered the electricity generated by the application
of heat to the junction of two different materials (1821, Seebeck
effect) \cite{seebeck21,seebeck26}, Jean C. A. Peltier found a temperature
gradient in the junction under isothermal conditions due an electrical
current (1834, Peltier effect) \cite{peltier34}, and W. Thomson predicted
and observed the heating or cooling of a current-carrying conductor
with a temperature gradient (1851, Thomson effect) \cite{callen85,tribus61}.
Thomson's experiments allowed him to find two relations between these
effects: one was a subtle connection between the Peltier effect and
the Sebeeck effect, called Second Thomson's Relation (STR). The other
was a relation between the three effects, called First Thomson's Relation
(FTR). It was not until the advent of the linear theory of non\textendash equilibrium
processes, established by L. Onsager \cite{onsager31a,onsager31b},
that it was possible to satisfactorily demonstrate both relations.

L. Onsager first and later several authors \cite{callen85,onsager31a,onsager31b,garcia03,degroot84,callen48},
derived the phenomenological equations of the thermocouple. Begining
with the entropy production of thermoelectric phenomenon and considering
the electrochemical potential and the fluxes and forces on the system,
we obtain the generalized equations \cite{callen85}, 
\begin{equation}
\left[\begin{array}{c}
-J_{N}\\
J_{Q}
\end{array}\right]=\left[\begin{array}{cc}
L_{11} & L_{12}\\
L_{21} & L_{22}
\end{array}\right]\left[\begin{array}{c}
\frac{1}{T}\nabla\mu\\
\nabla\left(\frac{1}{T}\right)
\end{array}\right],\label{egtp}
\end{equation}
with $-J_{N}$ the electrical current (the generalized flux $J_{1}$),
$J_{Q}$ the heat flux (the generalized flux $J_{2}$), $L_{ij}'s$
the Onsager coefficients. For the Seebeck effect, we can take as the
generalized driven force the electric potential $X_{1}=\nabla\mu/eT$,
and take as the driver generalized force the temperature gradient
$X_{2}=-\nabla\left(1/T\right)$. These gradients are between the
welding points of materials $A$ and $B$ (see Figure \ref{fig1}a).
Then we get the phenomenological Onsager's equations:
\begin{equation}
\left[\begin{array}{c}
J_{1}\\
J_{2}
\end{array}\right]=\left[\begin{array}{cc}
L_{11} & L_{12}\\
L_{21} & L_{22}
\end{array}\right]\left[\begin{array}{c}
X_{1}\\
X_{2}
\end{array}\right].\label{egons}
\end{equation}
where $L_{ij}=\left(\frac{\partial J_{i}}{\partial X_{i}}\right)_{eq}$.
Now, from the entropy production of the thermocouple, 
\begin{equation}
\sigma=J_{1}X_{1}+J_{2}X_{2}>0\label{s}
\end{equation}

we can establish the relation, $\left|J_{2}\,X_{2}\right|>\left|J_{1}\,X_{1}\right|,$
with $J_{1}\,X_{1}<0$ and $J_{2}\,X_{2}>0$, agree with the definition
of the driven and driver forces respectively. Then, we can associated
the first term of the entropy production to a power output (by temperature
unit) and the second to a power input (by temperature unit), and build
a steady state Linear Energy Converter (LEC) \cite{caplan83,arias08}
(see Figure \ref{fig1}b). This array is a nonzero entropy production
and a nonzero power output converter, because of its interactions
with the surroundings ($X_{i}$ and $J_{i}$). Using the work of Caplan
and Essig \cite{caplan83}, it is possible to make a first step towards
a linear description of a LEC. In general, the governing fluxes $J_{i}$
of a real system are usually very complicated and non\textemdash linear
functions of the generalezd forces $X_{i}$. However, the linear regime
allows us to give a fair enough description of the phenomenon. These
authors, based on the analysis of equations (\ref{egons}) introduced
the so called coupling coefficient $q\left(L_{ij}\right)$, which
comes directly from the second law of thermodynamics \cite{arias08}.
This is a dimensionless parameter that measures the degree of coupling
between the spontaneous and nonspontaneous fluxes, 
\begin{equation}
1\geq q^{2}\equiv\frac{L_{12}^{2}}{L_{11}L_{22}}\geq0.\label{q}
\end{equation}
In addition, we can take into account a parameter introduced by Stucki
\cite{stucki80} which measures the relation between the two forces
$X_{1}$ and $X_{2}$ as follows: 
\begin{equation}
x\equiv\sqrt{\frac{L_{11}}{L_{22}}}\frac{X_{1}}{X_{2}}\quad\frac{\longleftarrow driven\;force}{\longleftarrow Driver\;force},\label{x}
\end{equation}
where $x\in\left[-1,0\right]$ is called the force ratio; also we
can build $J_{1}/J_{2}=\sqrt{L_{11}/L_{22}}$ the ratio between the
fluxes.

On the other hand, one of the most important features of an irreversible
converter, is the amount of energy exchanged with the surroundings
to do work or acomplish another type of objective. This feature is
usually known as the energetics of the LEC \cite{arias08}. We can
write some functions that characterize this energetics, in terms of
the parameters $q$, $x$, $L_{22}$, and the force $X_{2}$.

This paper is organized as follows: in Section \ref{lec} we present
a non\textendash isothermic LEC and different working regimens of
this converter, transferred here from Finite\textendash Time Thermodynamics
(FTT) \cite{angulo91,arias97,angulo01,arias03,calvo01,paez,wagner,van05}.
The converter can be operated as a heat engine (direct energy converter)
or as a refrigerator (inverse energy converter). In Section \ref{rel}
we present the deduction of the phemomenological coefficients of thermocouple,
starting from the phenomenological equations and the general form
of its entropy production. Then we introduce the operation modes built
in Section \ref{lec} with the objective to rewrite the Thomson's
relations considering the thermocouple as a LEC. Finally, in \ref{conc}
we present some conclusions concerning our results.

\section{Non\textendash isothermic LEC optimization\label{lec}}

\begin{figure}[t]
\centering a) %
\fbox{\includegraphics[width=7cm,height=5cm]{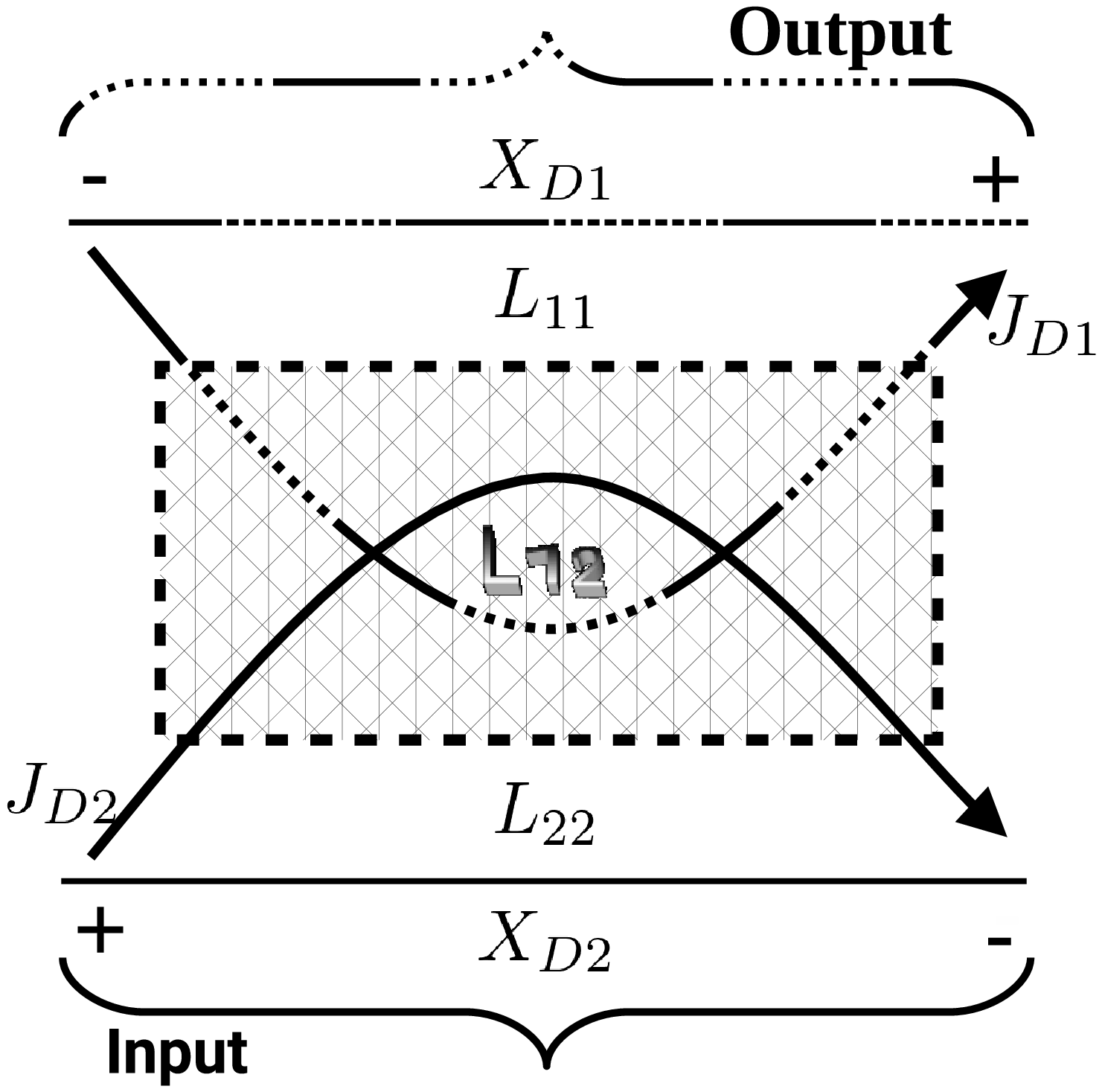} } \hspace*{0.5cm}b) %
\fbox{\includegraphics[width=7cm,height=5cm]{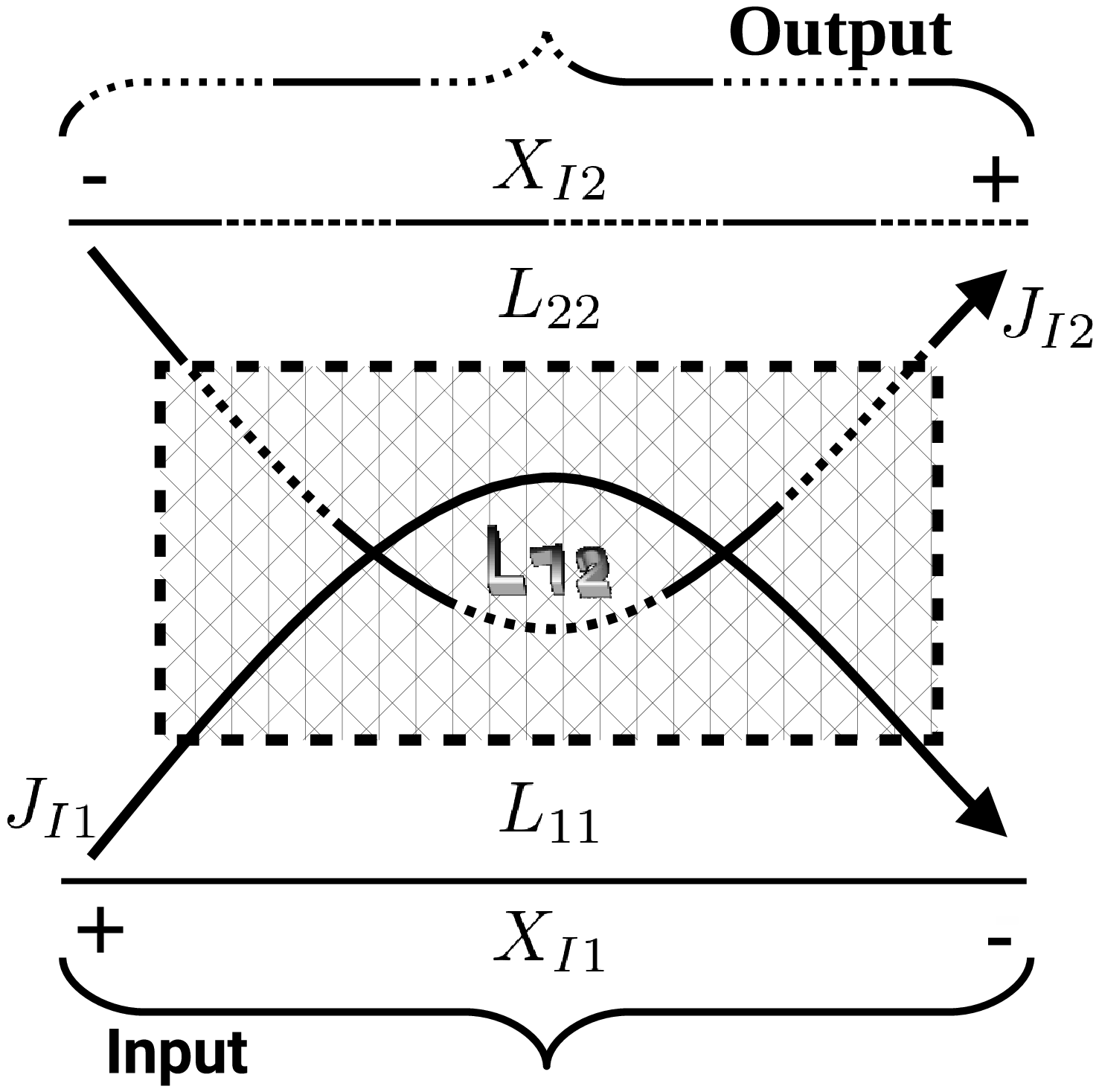} } \caption{Steady non-isothermic Linear Energy Converters. a) The heat engine
as a direct linear energy converter (D\textendash LEC). We can describe
this engine with the scheme shown here, a system with two fluxes ($J_{D1}$,
$J_{D2}$) and two forces ($X_{D1}$, $X_{D2}$), where $J_{D2}$
is the input heat flux, then $\left|J_{D2}\,X_{D2}\right|$ is the
power input (by temperature unit) and $\left|J_{D1}\,X_{D1}\right|$
is the power output (by temperature unit) of the converter. b) The
refrigerator as an inverse linear energy converter (I\textendash LEC).
In this case we have two fluxes ($J_{I1}$, $J_{I2}$) and two forces
($X_{I1}$, $X_{I2}$), but now $\left|J_{D2}\,X_{D2}\right|$ is
the power output (by temperature unit) and $\left|J_{D1}\,X_{D1}\right|$
is the power input (by temperature unit) of the refrigerator.}
\label{fig1} 
\end{figure}

Of all actual energy converters, a very large portion of them use
gradients of temperature. A set of these thermal engines are converters
as the thermocouple and other systems which contains pairs of fluxes
that give us cross\textendash effects, such as the Soret effect or
Reynolds effect \cite{kohler}. With the purpose of make a general
study of the energetics of these kind of phenomena, in the next paragraphs
we will take the entropy production in two cases: when the heat flux
is a spontaneous flux (see Fig. \ref{fig1}a) and when this flux is
non-spontaneous (see Fig. \ref{fig1}b). We call the first case direct
converter and the second case inverse converter. Later we will use
some known objective functions of models of irreversible energy converters
studied in other contexts, and built the equivalent objective functions
for these new models. 

\subsection{Heat engine (direct LEC)}

One of the most common thermal engines, is that exchanging an amount
of energy with the surroundings to do work, known as a heat engine.
In this case a gradient of temperature promotes a flux against any
other gradient (gravity, electric field, etc.). Some models of this
kind of engines have been proposed in the context of Linear Irreversible
Thermodynamics, Finite Time Thermodynamics and other constructions
within Non\textendash Equilibrium Thermodynamics \cite{van05,arias08}.

Now we can use the entropy production of the LEC, given in general
form by eq. (\ref{s}), and take as the driver flux the heat flux,
and as the driven flux any other flux against a generalized force.
For this reason we will call this engine ``direct linear energy converter''
(D\textendash LEC). In this case the force ratio will be
\begin{equation}
x_{D}=\sqrt{\frac{L_{11}}{L_{22}}}\frac{X_{D1}}{X_{D2}}.\label{xd}
\end{equation}
Now, using the $q$ and $x_{D}$ parameters we can write the flows
$J_{D1}$ and $J_{D2}$ as follows,

\begin{equation}
J_{D1}\left(x_{D},q\right)=\begin{cases}
\left(1+\frac{q}{x_{D}}\right)L_{11}X_{D1} & \quad a)\\
\left(1+\frac{x_{D}}{q}\right)L_{12}X_{D2} & \quad b)
\end{cases},\label{jd1}
\end{equation}
and

\begin{equation}
J_{D2}\left(x_{D},q\right)=\begin{cases}
\left(1+\frac{1}{qx_{D}}\right)L_{12}X_{D1}\quad & a)\\
\left(qx_{D}+1\right)L_{22}X_{D2}\quad & b)
\end{cases}.\label{jd2}
\end{equation}

We note that both \ref{jd1}a and \ref{jd1}b as \ref{jd2}a and \ref{jd2}b
are equivalent since it can be reached from one to another by performing
the proper substitution of the force ratio and the coupling parameter.
Now, we make an additional hypothesis about the driver force; we will
suppose that the temperature gradient is constant and of the form,
\begin{equation}
X_{D2}=\frac{1}{T_{c}}-\frac{1}{T_{h}}>0,\label{x2}
\end{equation}
with $T_{c}$ the temperature of the ``cold'' reservoir and $T_{h}$
the temperature of the ``hot'' reservoir. Due to this hypothesis
the D\textendash LEC is a steady state converter.

\subsubsection{D\textendash LEC Dissipation $\Phi_{D}^{*}$}

\begin{figure}[tb]
\begin{centering}
a) \includegraphics[width=7cm,height=5cm]{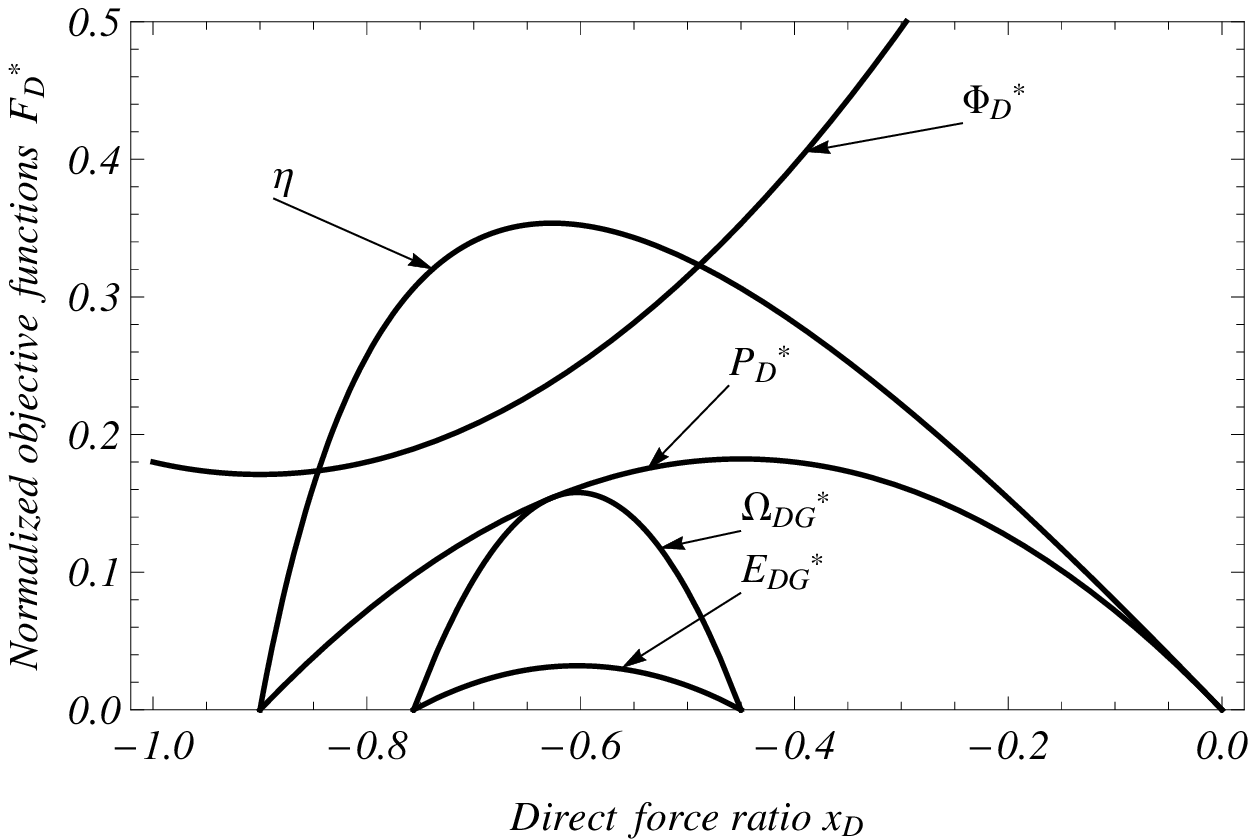}\hspace*{0.35cm}b)
\includegraphics[width=7cm,height=5cm]{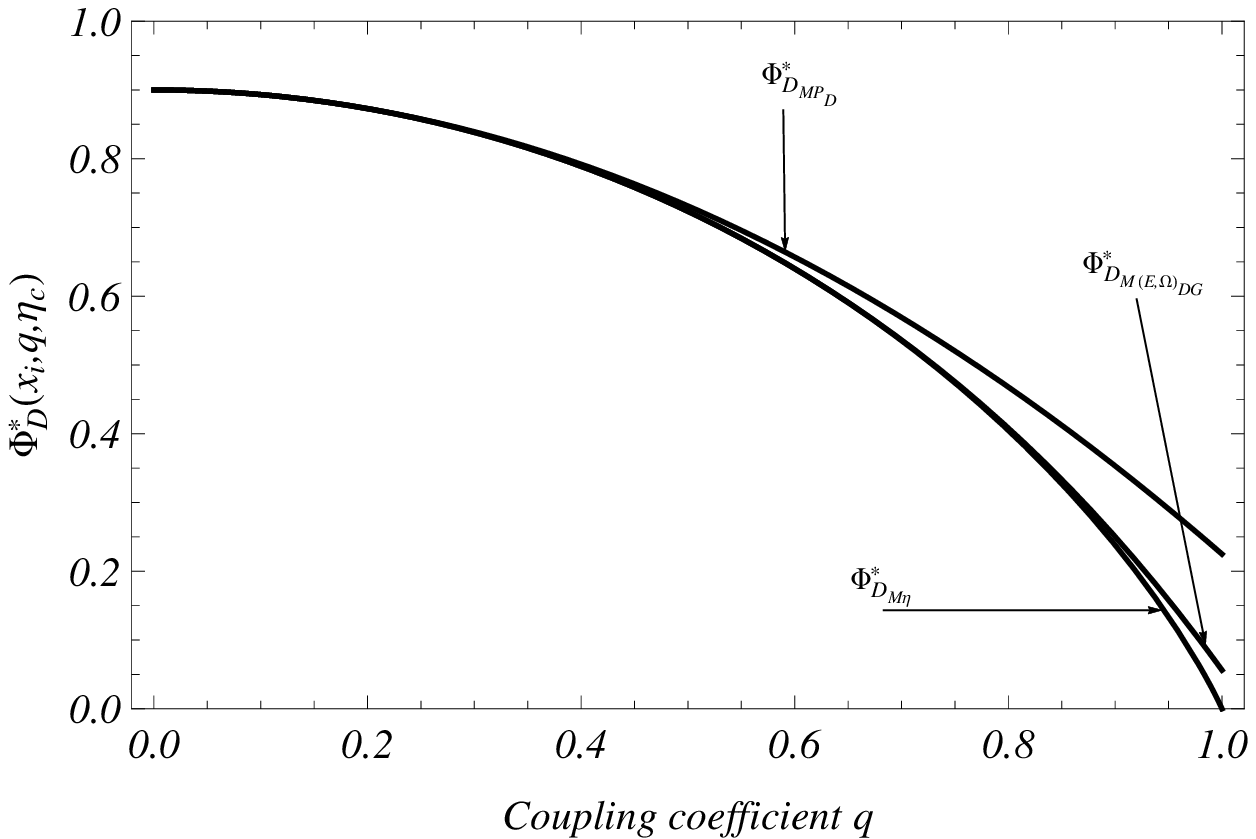}
\par\end{centering}
\begin{centering}
c) \includegraphics[width=7cm,height=5cm]{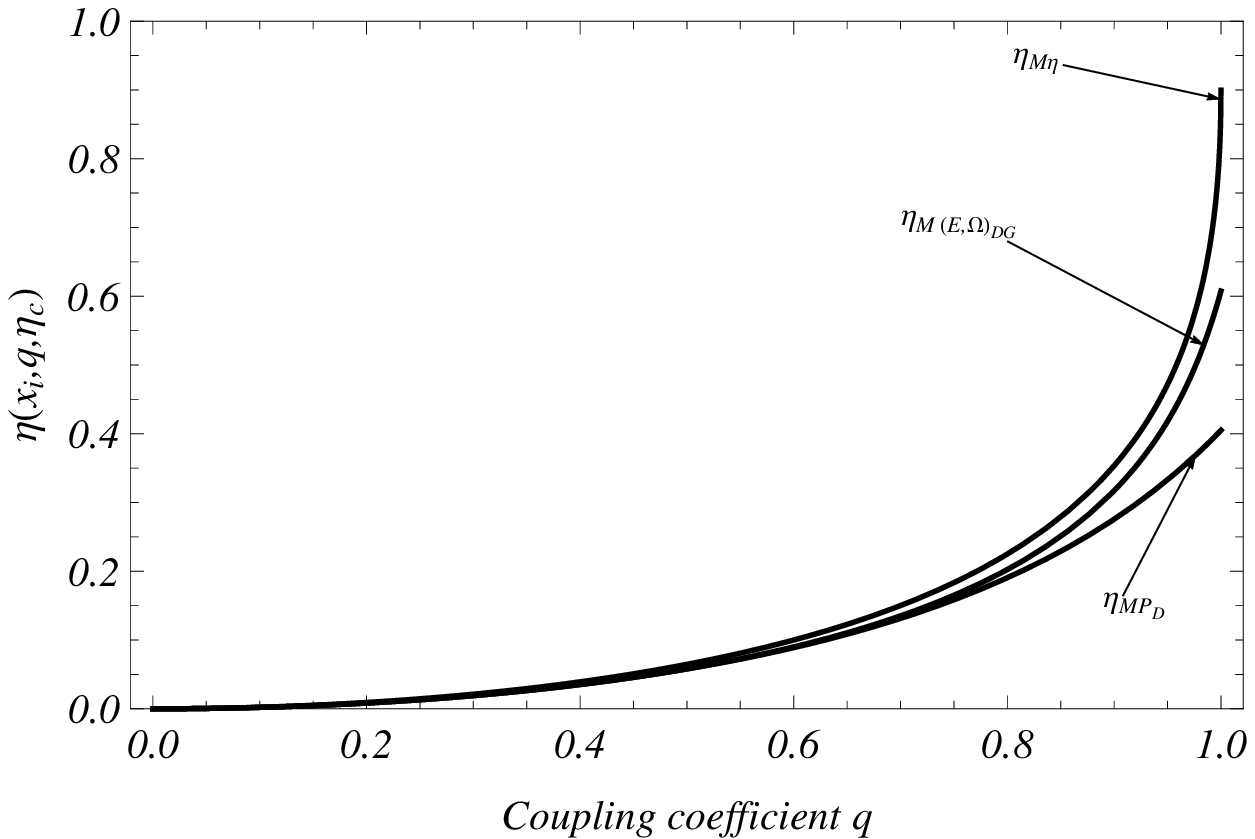}\hspace*{0.35cm}d)
\includegraphics[width=7cm,height=5cm]{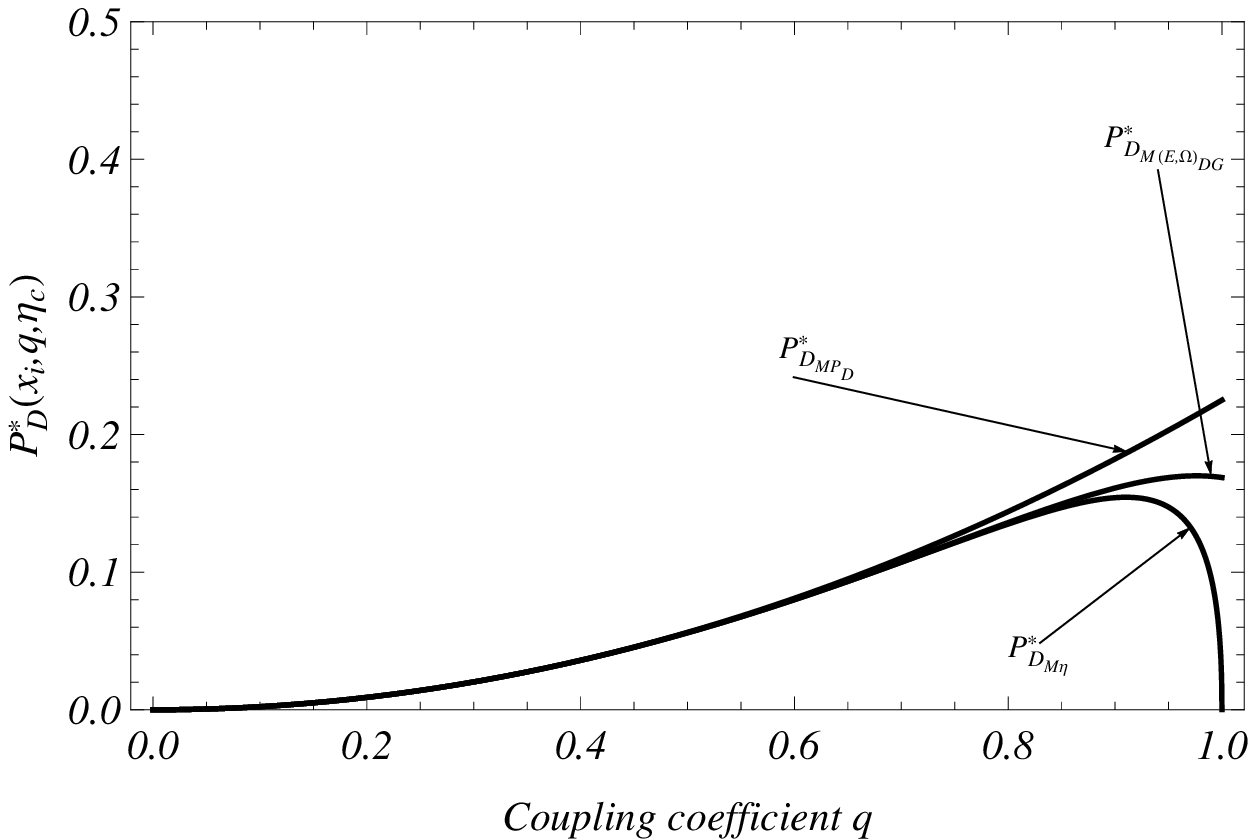}
\par\end{centering}
\begin{quote}
\caption{a) Different objective functions for the steady state non\textendash isothermic
D\textendash LEC: Dissipation function $\Phi_{D}^{*}$, Efficiency
$\eta$ (not normalized), Power output $P_{D}^{*}$, Generalized ecological
function $E_{DG}^{*}$, Generalized omega function $\Omega_{DG}^{*}$.
Here we take $q=0.9$ and $\eta_{C}=0.9$. b) Comparative plot of
the dissipation function at different working regimes, note $\Phi_{D}^{*}\left(x_{mdf},q\right)=\Phi_{D_{mdf}}^{*}=\xrightarrow{q\rightarrow1}0$.
c) Comparative plot of the efficiency at different working regimes,
note $\eta\left(x_{mdf},q\right)=\eta_{mdf}=\xrightarrow[x\rightarrow-q]{q\rightarrow1}\eta_{C}$.
d) Comparative plot of the power output at different working regimes,
note $P_{D}^{*}\left(x_{mdf},q\right)=P_{D_{mdf}}^{*}=0$ }

\label{fig2}
\end{quote}
\end{figure}

On the basis of the analysis of the entropy production (Eq. \ref{s})
it is possible to construct several objective functions for this D\textendash LEC.
The first function that we can construct is a function called dissipation
($\Phi_{D}$). At first approximation this function can be considered
as measuring the part of energy that is used only for the coupling
between the driver and the driven flux. We define $\Phi_{D}$ in terms
of generalized forces and fluxes through the entropy production as
follows \cite{tribus61,odum55},

\begin{equation}
\Phi_{D}\equiv T_{c}\left(J_{D1}X_{D1}+J_{D2}X_{D2}\right)=T_{c}X_{D2}\left(J_{D1}\frac{X_{D1}}{X_{D2}}+J_{D2}\right)=\eta_{C}\left(x_{D}^{2}+2x_{D}q+1\right)L_{22}X_{D2},\label{disd}
\end{equation}
here we use the explicit form of the temperature gradient to obtain
$\eta_{C}=T_{c}X_{D2}=\left(1-T_{c}/T_{h}\right)$, substitute the
Eqs. (\ref{jd1}b) and (\ref{jd2}b) in Eq. (\ref{s}), and we get
Eq. (\ref{disd}) in terms of $\eta_{C}$, $x$, $q$, $L_{22}$ and
$X_{D2}$. Finally we normalize the dissipation function by the constat
$L_{22}X_{D2}$:

\begin{equation}
\Phi_{D}^{*}\left(x_{D},q,\eta_{C}\right)=\frac{\Phi_{D}}{L_{22}X_{D2}}=\left(x_{D}^{2}+2qx_{D}+1\right)\eta_{C}.\label{disdn}
\end{equation}
This expression is analog to that published by Arias\textendash Hernandez
et al for a steady state isothermic\textendash LEC (see ec. 8 of \cite{arias08}).
Hereinafter we will consider normalized functions such that $F^{*}=F/L_{22}X_{D2}$.
The nomalized dissipation $\Phi_{D}^{*}$ is plotted versus the force
ratio $x$ in Fig. \ref{fig2}a, in this graphic we observe that $\Phi_{D}^{*}$
has a minimum.
 We can optimize the D\textendash LEC, with the purpose that it operates
in a working regime of minimum dissipation ($mdf$), by finding the
value of the force ratio $x_{mdf}$ which satisfies the equation $\left.\partial_{x_{D}}\Phi_{D}^{*}\left(x_{D},q,\eta_{C}\right)\right|_{x_{mdf}}=0$,

\begin{equation}
x_{mdf}\left(q\right)=-q.\label{xmdf}
\end{equation}

\subsubsection{D\textendash LEC Power output $P_{D}^{*}$}

Another objective function that we could built is the power output
of the D\textendash LEC. From the dissipation function (Eq. \ref{s})
we note that the first term $T_{c}J_{D1}X_{D1}<0$, which corresponds
to the driven flux promoted against a generalized gradient, has units
of energy per second, which can be taken as the power output $P\equiv-T_{c}J_{D1}X_{D1}$
of the D\textendash LEC. Now if we take \ref{x} and \ref{jd1}b and
replace them in $P$ we obtain, 
\begin{equation}
P_{D}^{*}\left(x_{D},q,\eta_{C}\right)=-x_{D}\left(x_{D}+q\right)\eta_{C}.\label{pd}
\end{equation}
This function is plotted in Figure \ref{fig2}a and we can observe
that it has a maximum, so there exists a $x_{MP_{D}}$ solution of
$\left.\partial_{x_{D}}P_{D}^{*}\left(x_{D},q,\eta_{C}\right)\right|_{x_{MP_{D}}}=0$
and a maximum power output working regime ($MP_{D}$) is possible
to operate the D\textendash LEC, if 
\begin{equation}
x_{MP_{D}}\left(q\right)=-\frac{q}{2}.\label{xmpd}
\end{equation}

\subsubsection{D\textendash LEC Efficiency $\eta$}

We can define the irreversible efficiency of the D\textendash LEC,
as the power output divided by the input heat flux $\eta\equiv P/J_{2}$,
and using Eqs. (\ref{pd}) and (\ref{jd2}b) we get,

\begin{equation}
\eta\left(x_{D},q,\eta_{C}\right)=-\eta_{C}\frac{x_{D}\left(x_{D}+q\right)}{1+qx_{D}}.\label{efd}
\end{equation}
Note that the efficiency is not a function of $L_{22}X_{D2}$. We
plot $\eta$ versus $x_{D}$ and see in Figure \ref{fig2} that it
has a maximum. This maximum is given by, 

\begin{equation}
x_{M\eta}\left(q\right)=-\frac{q}{1+\sqrt{1-q^{2}}},\label{xmef}
\end{equation}
obtained from the equation $\left.\partial_{x_{D}}\eta\left(x_{D},q,\eta_{C}\right)\right|_{x_{M\eta}}=0$.
Therefore, the D\textendash LEC can operate in an optimum efficiency
working regime ($M\eta$).

\subsubsection{D\textendash LEC Generalized ecological function $E_{DG}^{*}$}

Using the characteristic functions we can built functions that accomplish
other objectives, for example a good trade\textendash off between
the dissipation and the power output. Within the context of Finite
Time Thermodynamics (FTT), in 1991, F. Angulo\textendash Brown \cite{angulo91}
proposed the Ecological Function, $E=P_{D}-\Phi_{D}$, as this good
trade\textendash off function. If we operate the heat engine at maximum
ecological working regime, the engine reaches around 80\% of the power
output of the $MP_{D}$\textendash working regime and 30\% of the
dissipation of this regime \cite{arias97}. Later the Generalized
Ecological Function was proposed \cite{angulo01,tornez06}, that guaranteed
the best trade\textendash off between the power output and dissipation,
through the function $g_{D}^{E}\left(\eta\right)=\eta/\left(\eta_{C}-\eta\right)$
\cite{angulo91,arias97,angulo01,arias03} evaluated at the efficiency
of the $MP_{D}$\textendash working regime, $E_{DG}=P_{D}-g_{MP_{D}}^{E}\Phi_{D}$.
Evaluating the efficiency (Eq. \ref{efd}) at $x_{MP_{D}}=-q/2$ we
get:
\begin{equation}
\eta\left(x_{MP_{D}},q,\eta_{C}\right)=\eta_{MP_{D}}\left(\eta_{C},q\right)=\frac{q^{2}}{2\left(2-q^{2}\right)}\eta_{C},\label{empd}
\end{equation}
and substituting in $g$,

\begin{equation}
g_{D}^{E}\left[\eta_{MP_{D}}\left(\eta_{C},q\right)\right]=g_{MP_{D}}^{E}\left(q\right)=\frac{q^{2}}{4-3q^{2}}.\label{gmpd}
\end{equation}
Note that in the limit of ideal coupling we have $\lim_{q\rightarrow1}g_{MP_{D}}^{E}\left(q\right)=1$
and $E_{DG}=E$. 

Finally, using Eqs. (\ref{pd}), (\ref{disdn}) and (\ref{gmpd})
we write the generalized ecological function for the D\textendash LEC
as,

\begin{equation}
E_{DG}^{*}\left(x_{D},q,\eta_{C}\right)=\frac{x_{D}\left[4x_{D}-q\left(q^{2}-4\right)\right]+q^{2}\left(1-2x_{D}^{2}\right)}{3q^{2}-4}\eta_{C}.\label{ecd}
\end{equation}
We show the plot of $E_{DG}$ versus $x$, for a given $q$ and $\eta_{C}$,
in Figure \ref{fig2}a and observe that this function has a maximum.
Then the generalized ecological function can be used to optimize the
operation D\textendash LEC at this point ($x_{ME_{DG}}$); we call
this the $ME_{DG}$\textendash working regime. Solving the equation
$\left.\partial_{x_{D}}E_{DG}^{*}\left(x_{D},q,\eta_{C}\right)\right|_{x_{ME_{DG}}}=0$
we obtain $x_{ME_{DG}}$,

\begin{equation}
x_{ME_{DG}}\left(q\right)=-\frac{q^{2}\left(q^{2}-4\right)}{4\left(q^{2}-2\right)}.\label{xmecd}
\end{equation}

\subsubsection{D\textendash LEC Generalized omega function $\Omega_{DG}^{*}$}

The last objective function that we built for the D\textendash LEC,
is the generalized omega function. In 2001 within the context of FTT,
Calvo\textendash Hernandez et al proposed a unified optimization criterion
for energy converters, based on the maximum of the omega function
$\Omega=E_{eu}-E_{lu}$ \cite{calvo01}. This function mades a trade\textendash off
between the effective useful energy $E_{eu}\equiv E_{u}-r_{min}E_{i}$,
and the lost useful energy $E_{lu}\equiv r_{max}E_{i}-E_{u}$ , where
$E_{u}$ is the useful energy of the heat engine, $r_{min}$ is its
minimum performance, $E_{i}$ is the input energy and $r_{max}$ is
its maximum performance. The performance of the engine is defined
as $r\equiv E_{u}/E_{i}$. The generalized function $\Omega_{DG}=E_{eu}-g_{MP_{D}}^{\Omega}E_{lu}$,
was introduced by Tornez in 2006 \cite{tornez06}, where $g_{MP_{D}}^{\Omega}$
is the function $g_{D}^{\Omega}\left(\eta\right)=\eta/\left(\eta_{M\eta}-\eta\right)$
for the omega function evaluated in the $MP_{D}$\textendash efficiency
(Eq. \ref{empd}),
\begin{equation}
g_{D}^{\Omega}\left[\eta_{MP_{D}}\left(\eta_{C},q\right)\right]=g_{MP_{D}}^{\Omega}\left(q\right)=\left(\frac{\sqrt{1-q^{2}}+1}{\sqrt{1-q^{2}}-1}\right)^{2}.\label{gmpdo}
\end{equation}
Operating the engine in the $M\Omega_{DG}$\textendash working regime
achieves the best compromise between $E_{eu}$ and $E_{lu}$. In the
context of this model we can use the dissipation to define the input
energy of the D\textendash LEC: $E_{i}\equiv TJ_{2}X_{2}=\eta_{C}\left(qx+1\right)L_{22}X_{2}$.
On the other hand its useful energy is the power output $E_{u}\equiv P_{D}$,
so the performance for the D\textendash LEC is,
\begin{equation}
r_{D}\left(x_{D},q\right)=-\frac{x_{D}\left(x_{D}+q\right)}{\left(qx_{D}+1\right)},\label{rd}
\end{equation}
which is related to the efficiency in the following manner $r_{D}=\eta/\eta_{C}$,
from this relation we conclude that the minimum performance of the
D\textendash LEC is $r_{min}=0$, and the maximum is $r_{max}=\eta_{M\eta}/\eta_{C}$,
where $\eta_{M\eta}$ is the efficiency evaluted at $x_{M\eta}$ (Eq.
\ref{xmef}) and is given by, 
\begin{equation}
\eta\left(x_{M\eta},q,\eta_{C}\right)=\eta_{M\eta}\left(\eta_{C},q\right)=\eta_{C}\left(\frac{q}{1+\sqrt{1-q^{2}}}\right)^{2}\label{eme}
\end{equation}
Substituting $E_{u}$, $E_{i}$, $r_{min},$ $r_{max}$ and $g_{MP_{D}}^{\Omega}$
in the definitions of $E_{eu}$, $E_{lu}$and $\Omega_{DG}$ we obtain,

\begin{equation}
\Omega_{DG}^{*}\left(x_{D},q,\eta_{C}\right)=\frac{x_{D}\left[q\left(q^{2}-4\right)-4x_{D}\right]+q^{2}\left(2x_{D}^{2}-1\right)}{\left(\sqrt{1-q^{2}}-1\right)^{2}}\eta_{C}.\label{od}
\end{equation}
In Figure \ref{fig2} we show that $\Omega_{DG}^{*}$ reaches it maximum
at $x_{M\Omega_{DG}}$,
\begin{equation}
x_{M\Omega_{DG}}=-\frac{q^{2}\left(q^{2}-4\right)}{4\left(q^{2}-2\right)},\label{xmod}
\end{equation}
 which is the solution of $\left.\partial_{x_{D}}\Omega_{DG}^{*}\left(x_{D},q,\eta_{C}\right)\right|_{x_{M\Omega_{DG}}}=0$.

\subsubsection{Energetics of the D\textendash LEC}

The energetics of the D\textendash LEC is shown in Figure \ref{fig2}.
We must note that $x_{ME_{DG}}=x_{M\Omega_{DG}}$, therefore the characteristic
functions in these working regimes are the same. We show the characteristic
functions of the above working regimes in Table \ref{t1} {\tiny{}}
\begin{table}[tb]
{\tiny{}\centering}%
\begin{tabular}{cccccc}
\toprule 
\multirow{1}{*}{D\textendash LEC} &  & $x_{mdf}=-q$ & $x_{M\eta}=-\frac{q}{1+\sqrt{1-q^{2}}}$ & $x_{MP_{D}}=-\frac{q}{2}$ & $x_{M\left(E,\Omega\right)_{DG}}=-\frac{q^{2}\left(q^{2}-4\right)}{4\left(q^{2}-2\right)}$\tabularnewline
working regimes &  &  &  &  & \tabularnewline
\midrule 
 & $\Phi_{D}^{*}(x_{D},q)=$  & $\xrightarrow{q\rightarrow1}0$  & $\eta_{C}\frac{2\left(1-q^{2}\right)}{1+\sqrt{1-q^{2}}}$ & $\eta_{C}\left(1-\frac{3q^{2}}{4}\right)$ & $\eta_{C}\frac{64-7\left(q^{3}-4q\right)^{2}}{16\left(q^{2}-2\right)^{2}}$\tabularnewline
\midrule 
 & $\eta(x_{D},q)=$  & $\eta_{mdf}\xrightarrow[x\rightarrow-q]{q\rightarrow1}\eta_{C}$ & $\eta_{C}\left(\frac{q}{1+\sqrt{1-q^{2}}}\right)^{2}$  & $-\eta_{C}\frac{q^{2}}{2\left(q^{2}-2\right)}$ & $-\eta_{C}^{2}\frac{q^{2}\left(q^{2}-4\right)\left(3q^{2}-4\right)}{32\left(q^{2}-2\right)\left(1-q^{2}+q^{4}\right)}$\tabularnewline
\midrule 
 & $P_{D}^{*}(x_{D},q)=$  & $0$ & $\eta_{C}\sqrt{1-q^{2}}\left(\frac{q}{1+\sqrt{1-q^{2}}}\right)^{2}$ & $\eta_{C}\left(\frac{q}{2}\right)^{2}$  & $\eta_{C}\frac{q^{2}\left(q^{2}-4\right)\left(3q^{2}-4\right)}{16\left(q^{2}-2\right)^{2}}$\tabularnewline
\bottomrule
\end{tabular}{\tiny{} \caption{The table shows the characteristic functions for different {\small{}working
regimes} of the steady state non\textendash isothermic D\textendash LEC,
which are: Dissipation function $\Phi_{D}^{*}$, efficiency $\eta$
and Power output $P_{D}^{*}$, evaluated at the optimal values of
the force ratio of these working regimes, which are respectively,
minimum function of dissipation $x_{mdf}$, maximum efficiency $x_{M\eta}$,
maximum output power $x_{MP_{D}}$. Added to these optimal points
we include the force ratio of the maximum generalized ecological and
maximum generalized omega regime $x_{M\left(E,\Omega\right)_{DG}}$.}
}{\tiny \par}

{\tiny{}\label{t1} }{\tiny \par}
\end{table}
{\tiny \par}

If we observe the curves of the characteristic functions from Figure
\ref{fig2}, we will see that each of them represents a mode of operation
that fulfills some objective of the thermodynamic process, and that
the condition to operate the D-LEC optimally, to meet this objective,
is to achieve the corresponding force ratio $x_{i}$, that is, the
way in which the flow handled through its associated potential is
generated, and the potential against which the handler flux does work,
which is subject to a certain degree of fixed coupling given by the
design of the converter.

Based on the criteria analyzed here $\varPhi_{D}$, $P_{D}$, $\eta$
and $\left(E,\Omega\right)_{DG}$, we search for quotient ratios compatible
with the different optimization criteria (see Figs. \ref{fig2}).
For the D-LEC we observe the criteria comparatively ($mdf$, $M\eta$,
$MP_{D}$ and $M\left(E,\Omega\right)_{DG}$). Note that the efficiency
of a non-isothermal linear energy converter, working in the different
regimes saves the following hierarchy $\eta_{C}>\eta_{M\eta}>\eta_{M\left(E,\Omega\right)_{DG}}>\eta_{MP_{D}}$
(Fig. \ref{fig2}c). In the same way we can hierarchize the output
power of this converter operating in different modes (Fig. \ref{fig2}d),
such that $P_{DM\left(E,\Omega\right)_{DG}}^{*}>P_{DMP_{D}}^{*}>P_{DM\eta}^{*}>P_{Dmdf}^{*}$.
The dissipation of this converter $\varPhi_{D}^{*}$ evaluated in
the different working regimes satisfies the following hierarchy (Fig.
\ref{fig2}b), $\varPhi_{DMP_{D}}^{*}>\varPhi_{DM\left(E,\Omega\right)_{DG}}^{*}>\varPhi_{DM\eta}^{*}>\varPhi_{Dmdf}^{*}$.
Finally, note that in the limit of strong coupling, the force ratios
corresponding to the operating modes of the D-LEC, comply with the
following hierarchical order $x_{mdf}=x_{M\eta}<x_{M\left(E,\Omega\right)_{DG}}<x_{MP_{D}}$
(tab. \ref{t1}).

\subsection{Refrigerator (inverse LEC)}

About 30\% of world's energy is used to promote heat fluxes against
temperature gradients; these processes can be called inverse conversion
of energy. In particular, when the objective of the engine is to extract
a heat flux from a body, we could say that we have a refrigerator.

Now if we want to use the force ratio (Eq. \ref{x}) introduced by
Stucki \cite{stucki80} it is necessary to write it for the case when
the system is operating in an inverse mode, since that in the refrigerators
the driven flux is $J_{I2}=\dot{Q_{cI}}$ and the driver flux will
be $J_{I1}$ which are associated with the driven and driver forces
$X_{I2}$ and $X_{I1}$ respectively. Then the force ratio for the
I\textendash LEC is,

\begin{equation}
x_{I}=\sqrt{\frac{L_{22}}{L_{11}}}\frac{X_{I2}}{X_{I1}}.\label{xi}
\end{equation}
Now we can write the fluxs $J_{I1}$ and $J_{I2}$ in terms of the
inverse force ratio and the coupling coefficient as, 

\begin{equation}
J_{I1}=\begin{cases}
\left(1+qx_{I}\right)L_{11}X_{I1} & \quad a)\\
\left(1+\frac{1}{qx_{I}}\right)L_{12}X_{I2} & \quad b)
\end{cases},\label{ji1}
\end{equation}

and

\begin{equation}
J_{I2}=\begin{cases}
\left(1+\frac{x_{I}}{q}\right)L_{12}X_{I1}\quad & a)\\
\left(1+\frac{q}{x_{I}}\right)L_{22}X_{I2}\quad & b)
\end{cases}.\label{ji2}
\end{equation}

We will then extend the proposal of Jiménez de Cisneros et al \cite{jimenez06}
for the refrigeration cycles. We take as the driven force the following
force,
\begin{equation}
X_{I2}=\frac{1}{T_{h}}-\frac{1}{T_{c}}.\label{xi2}
\end{equation}
where $T_{h}$ and $T_{c}$ have the same meaning as in the D\textendash LEC.

\subsubsection{I\textendash LEC Dissipation $\Phi_{I}^{\#}$}

\begin{figure}[tb]
\begin{centering}
a) \includegraphics[width=7cm,height=5cm]{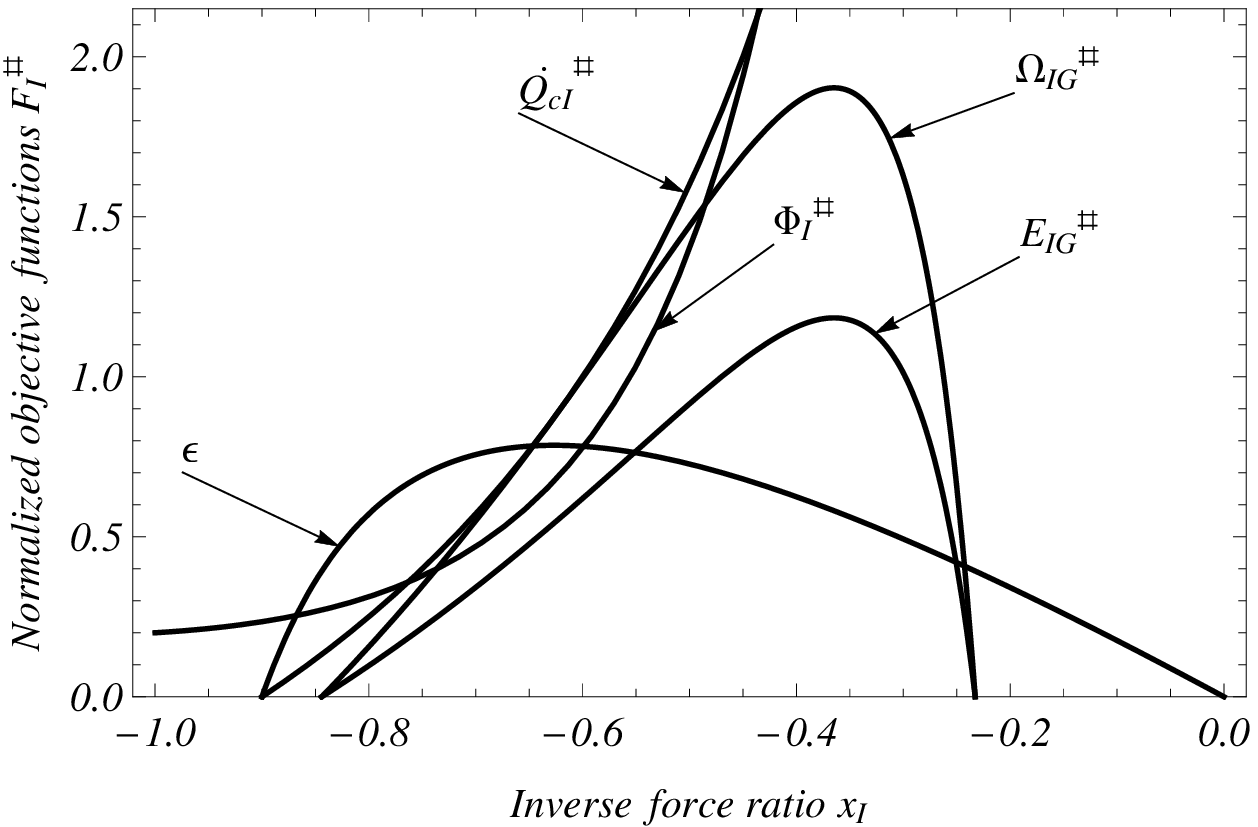}\hspace*{0.35cm}b)
\includegraphics[width=7cm,height=5cm]{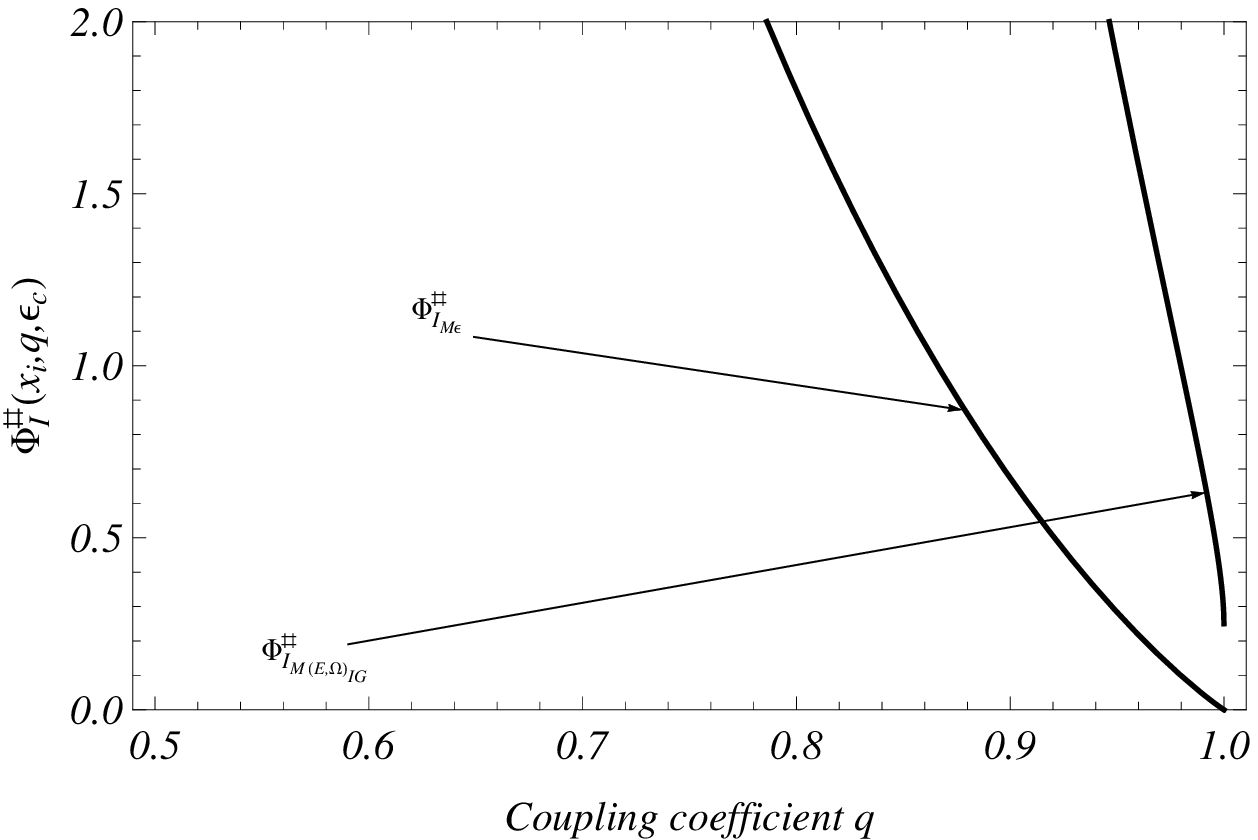}
\par\end{centering}
\begin{centering}
c) \includegraphics[width=7cm,height=5cm]{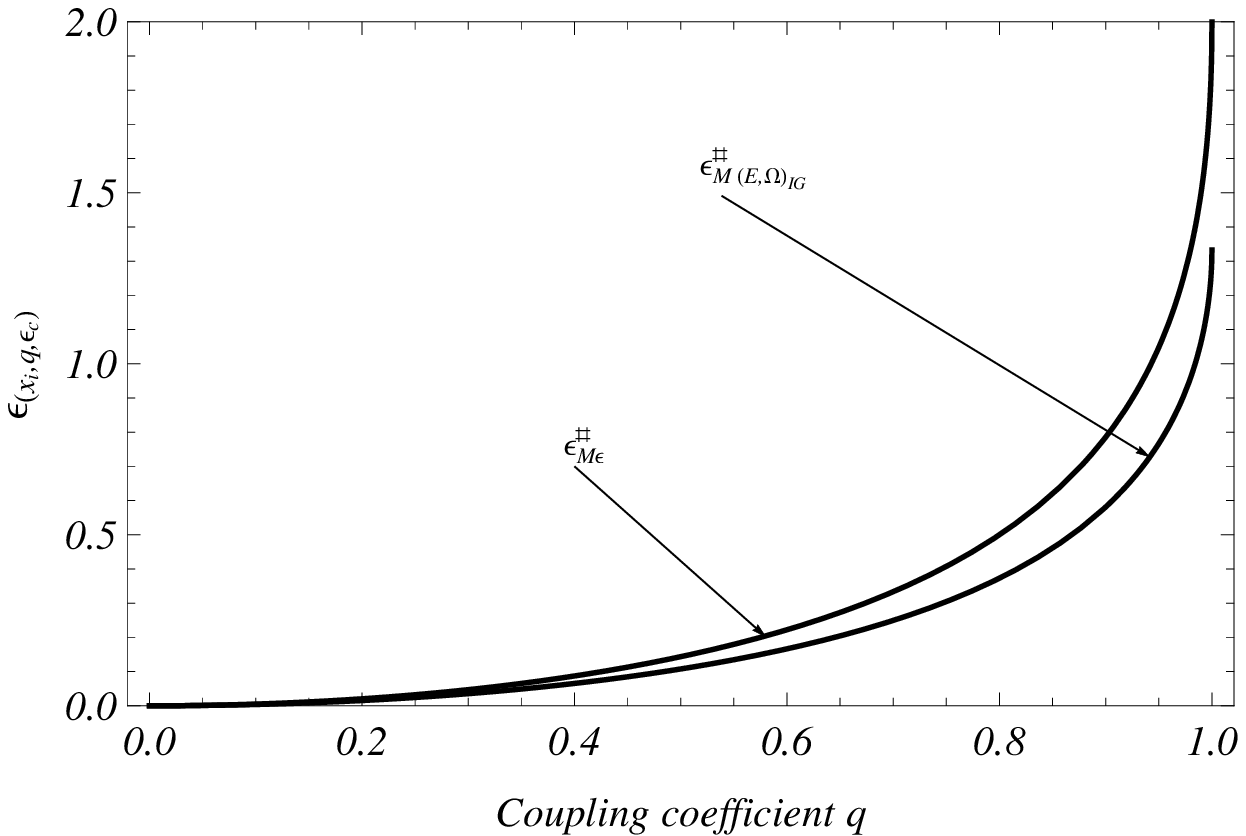}\hspace*{0.35cm}d)
\includegraphics[width=7cm,height=5cm]{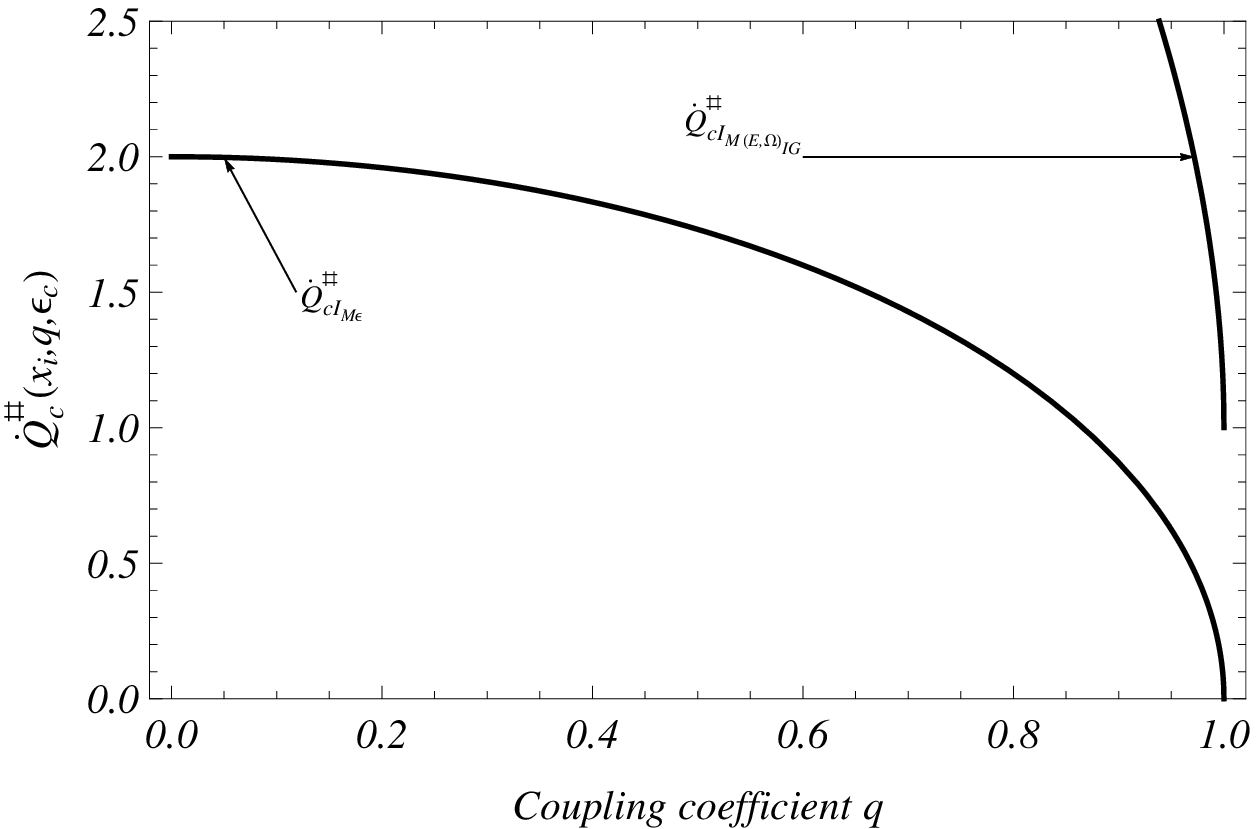}
\par\end{centering}
\begin{quote}
\caption{Different objective functions for the steady state non\textendash isothermic
I\textendash LEC: Dissipation function $\Phi_{I}^{\#}$, Coefficient
of performance $\epsilon$ (not normalized), Cooling power $\dot{Q}_{cI}^{\#}$,
Generalized ecological function $E_{IG}^{\#}$, Generalized omega
function $\Omega_{IG}^{\#}$. Here we take $q=0.9$ and $\epsilon_{C}=2$.
b) Comparative plot of the dissipation function at different working
regimes, note $\Phi_{I}^{\#}\left(x_{I_{mdf}},q\right)=\Phi_{I_{mdf}}^{\#}=\xrightarrow{q\rightarrow1}0$.
c) Comparative plot of the Coefficient of performance at different
working regimes, note $\epsilon\left(x_{I_{mdf}},q\right)=\epsilon_{I_{mdf}}=\xrightarrow[x\rightarrow-q]{q\rightarrow1}\epsilon_{C}$.
d) Comparative plot of the Cooling power at different working regimes,
note $J_{I2}=\dot{Q}_{cI}^{\#}(x_{I_{mdf}},q)=\dot{Q}_{cI_{mdf}}^{\#}=\xrightarrow{q\rightarrow1}0$. }

{\footnotesize{}\label{fig3} }{\footnotesize \par}
\end{quote}
\end{figure}

The dissipation function for the I\textendash LEC that operates between
these two reservoirs can be defined as $\Phi_{I}=T_{h}\sigma$, using
the entropy production Eq. (\ref{s}), substituting $x_{I}$, q we
get,

\begin{equation}
\Phi_{I}=T_{h}\left(J_{I1}X_{I1}+J_{I2}X_{I2}\right)=\left(T_{h}X_{I2}\right)\left(\frac{x_{I}^{2}+2qx_{I}+1}{x_{I}^{2}}\right)L_{22}X_{I2}=-\frac{1}{\epsilon_{C}}\left(\frac{x_{I}^{2}+2qx_{I}+1}{x_{I}^{2}}\right)L_{22}X_{I2},\label{di}
\end{equation}
and normalizing by $T_{h}L_{22}X_{I2}^{2}$ we obtain,

\begin{equation}
\Phi_{I}^{\#}\left(x_{I},q\right)=\frac{x_{I}^{2}+2qx_{I}+1}{x_{I}^{2}}.\label{din}
\end{equation}
The plot of this function versus $x_{I}$ for $q$ and $\epsilon_{C}$
fixed, shows a minimum in the Fig.\ref{fig3}a,
 and this minimum is reached at $x_{I_{mdf}}$ given by $\left.\partial_{x_{I}}\Phi_{I}^{\#}\left(x_{I},q\right)\right|_{x_{I_{mdf}}}=0$,

\begin{equation}
x_{I_{mdf}}\left(q\right)=-\frac{1}{q}.\label{ximdf}
\end{equation}
and the corresponding working regime of the I\textendash LEC ($mdf$\textendash working
regime), will be obtained when we evaluate its characteristic functions
in $x_{I_{mdf}}\left(q\right)=-1/q$.

\subsubsection{I\textendash LEC Coefficient of performance $\epsilon_{C}$}

As is well known the amount of heat flux that can be driven by a refrigerator
depends on the temperature difference between the reservoirs. The
greater the difference, the lower the engine performance. This performance
is measured by the Coefficient of Performance (COP) built with heat
flux extracted to the cold reservoir divided by the power input to
the I\textendash LEC, $\epsilon=\dot{Q_{cI}}/P$. In terms of generalized
fluxs and forces we can write,

\begin{equation}
\epsilon\left(x_{I},q,\epsilon_{C}\right)=\frac{J_{I2}}{T_{h}J_{I1}X_{I1}}=\frac{J_{I2}X_{I2}}{\left(T_{h}X_{I2}\right)J_{I1}X_{I1}}=-\epsilon_{C}\frac{x_{I}\left(x_{I}+q\right)}{1+qx_{I}},\label{cop}
\end{equation}
using the Eqs. (\ref{xi}), (\ref{ji1}a) and (\ref{ji2}b). From
Eq. (\ref{cop}) we see that $\epsilon$ has the same form of $\eta$,
but the values interval for $\epsilon_{C}$ is $\left[0,\infty\right)$.
COP is plotted in Fig.\ref{fig3}a and shows a maximum given by the
solution of $\left.\partial_{x_{I}}\epsilon\left(\epsilon_{C},q,x_{I}\right)\right|_{x_{M\epsilon}}=0$,

\begin{equation}
x_{M\epsilon}=-\frac{q}{1+\sqrt{1-q^{2}}},\label{xcop}
\end{equation}
we notice that $x_{M\eta}=x_{M\epsilon}$. At this point we get an
operating regime for the I\textendash LEC at maximum COP, the $M\epsilon$\textendash working
regime.

\subsubsection{I\textendash LEC Generalized ecological function $E_{IG}^{\#}$}

The generalized ecological function for an irreverible model of a
FTT\textendash refrigerator: $E_{G}^{R}=P_{e}-g_{M\epsilon_{MAX}}^{RE}T_{h}\sigma$,
was introduced by Tornez in 2006 \cite{tornez06}, it was defined
as a function whose objective is to obtain the best trade\textendash off
between the cooling power $P_{e}$ and the entropy production $T_{h}\sigma$
of the refrigerator, and the parameter $g^{RE}\left(\epsilon\right)=\epsilon_{C}\epsilon/\left(\epsilon_{C}-\epsilon\right)$
evaluated at half of the maximum COP, $g_{M\epsilon_{MAX}}^{RE}$,
guarantees this best trade\textendash off. We will define the generalized
ecological function for the I\textendash LEC as the difference between
the heat flux $J_{I2}=\dot{Q_{cI}}$ and the dissipation function
$\Phi_{I}$, in the following manner,

\begin{equation}
E_{IG}=\overset{.}{Q_{cI}}-g_{E}^{I}\left(\frac{\epsilon_{M\epsilon}}{2}\right)\Phi_{I},\label{ecig}
\end{equation}
where
\begin{equation}
\epsilon\left(x_{M\epsilon},q,\epsilon_{C}\right)=\epsilon_{M\epsilon}\left(\epsilon_{C},q\right)=\epsilon_{C}\left(\frac{q}{1+\sqrt{1-q^{2}}}\right)^{2},\label{cmc}
\end{equation}
then $g_{E}^{I}\left(\epsilon_{M\epsilon}/2\right)=\epsilon_{C}\left(\epsilon_{M\epsilon}/2\right)/\left[\epsilon_{C}-\left(\epsilon_{M\epsilon}/2\right)\right]$
will be,

\begin{equation}
g_{E}^{I}\left(\frac{\epsilon_{M\epsilon}}{2}\right)=\frac{\epsilon_{c}q^{2}}{2\left(1+\sqrt{1-q^{2}}\right)^{2}-q^{2}}.\label{gei}
\end{equation}

Substituting the generalized fluxes and forces (Eqs. \ref{xi}, \ref{ji1}
and \ref{ji2}) and Eq. (\ref{gei}) in Eq. (\ref{ecig}) and normalizing
by $T_{h}L_{22}X_{I2}^{2}$, we obtain the generalized ecological
function $E_{IG}^{\#}$ in terms of $x_{I}$, $q$ and $\epsilon_{C}$,

\begin{equation}
E_{IG}^{\#}\left(x_{I},q,\epsilon_{C}\right)=-\epsilon_{C}\left\{ 1+\frac{q}{x_{I}}\left[1+\frac{q}{x_{I}}\;\frac{x_{I}^{2}+2qx_{I}+1}{2\left(1+\sqrt{1-q^{2}}\right)^{2}-q^{2}}\right]\right\} .\label{ecigx}
\end{equation}
This function has a maximum (see Fig. \ref{fig3}a) therefore we can
operate the I\textendash LEC at this point and obtain a maximum $E_{IG}^{\#}$
working regime ($ME_{IG}$). To this end we take $\left.\partial_{x_{I}}E_{IG}^{\#}\left(x_{I},q,\epsilon_{C}\right)\right|_{x_{ME_{IG}}}=0$
and solve for $x_{ME_{IG}}$, 
\begin{equation}
x_{ME_{IG}}\left(q\right)=\frac{2q}{q^{2}-4\left(1+\sqrt{1-q^{2}}\right)},\label{xeig}
\end{equation}
then we can substitute this solution in the characteristic functions
of the I\textendash LEC to get the energetics of this working regime.

{\tiny{}}
\begin{table}[tb]
{\tiny{}\centering}%
\begin{tabular}{ccccc}
\toprule 
I\textendash LEC &  & $x_{I_{mdf}}=-\frac{1}{q}$ & $x_{M\epsilon}=x_{M\eta}$ & $x_{M\left(E,\Omega\right)_{IG}}=\frac{2q}{q^{2}-4\left(1+\sqrt{1-q^{2}}\right)}$\tabularnewline
working regimes &  &  &  & \tabularnewline
\midrule 
 & $\Phi_{I}^{\#}(x_{I},q)=$  & $\xrightarrow{q\rightarrow1}0$ & $\frac{2\left(1-q^{2}\right)\left(1+\sqrt{1-q^{2}}\right)}{q^{2}}$ & $\frac{5q^{2}}{4}+\frac{8\left(1+\sqrt{1-q^{2}}\right)}{q^{2}}-6\sqrt{1-q^{2}}-9$\tabularnewline
\midrule 
 & $\epsilon(x_{I},q)=$  & $\epsilon_{mdf}\xrightarrow[x\rightarrow-q]{q\rightarrow1}\epsilon_{C}$ & $\epsilon_{C}\left(\frac{q}{1+\sqrt{1-q^{2}}}\right)^{2}$ & $-\epsilon_{C}\frac{2q^{2}\left[q^{2}-2\left(1-2\sqrt{1-q^{2}}\right)\right]}{\left[4\left(1+\sqrt{1-q^{2}}\right)-3q^{2}\right]\left[4\left(1+\sqrt{1-q^{2}}\right)-q^{2}\right]}$\tabularnewline
\midrule 
 & $J_{I2}=\dot{Q}_{cI}^{\#}(x_{I},q)=$ & $\xrightarrow{q\rightarrow1}0$ & $\epsilon_{C}\sqrt{1-q^{2}}$ & $-\epsilon_{C}\left(1-\frac{q^{2}}{2}+2\sqrt{1-q^{2}}\right)$\tabularnewline
\bottomrule
\end{tabular}{\tiny{} \caption{The table shows the characteristic functions for different {\small{}working
regimes} of the steady state non\textendash isothermic I\textendash LEC,
which are: Dissipation function $\Phi_{I}^{*}$, Coefficient of performance
$\epsilon$ and Cooling power $\dot{Q}_{cI}^{\#}$, evaluated at the
optimal values of the force ratio of minimum function of dissipation
$x_{I_{mdf}}$, maximum coefficient of performance $x_{M\epsilon}$,
and the force ratio of the maximum generalized ecological and maximum
generalized omega regimes $x_{M\left(E,\Omega\right)_{IG}}$.}
}{\tiny \par}

{\tiny{}\label{t2} }{\tiny \par}
\end{table}
{\tiny \par}

\subsubsection{I\textendash LEC Generalized omega function $\Omega_{IG}^{\#}$}

The generalized omega function $\Omega_{G}^{R}=E_{eu}-g_{M\epsilon_{MAX}}^{R\Omega}E_{lu}$
was proposed by Tornez for an irreversible FTT\textendash refrigerator
\cite{tornez06}, the meaning of $E_{eu}$ and $E_{lu}$ are the same
as in the case of $\Omega$, but with the performance of the refrigerator
given by $r\equiv\left(P_{e}/P\right)=\epsilon$, with $P_{e}$ the
cooling power (useful energy, $E_{u}$) and $P$ the power supplied
(input energy, $E_{i}$). This objective function proposes a trade\textendash off
between the effective useful energy $E_{eu}$ and the lost useful
energy $E_{lu}$. The parameter $g_{M\epsilon_{MAX}}^{R\Omega}$ corresponds
to the function $g_{\Omega}^{R}\left(\epsilon\right)=\epsilon/\left(\epsilon_{MAX}-\epsilon\right)$
\cite{tornez06} evaluated at the half of maximum COP, therefore $g_{\Omega}^{R}\left(\epsilon_{MAX}/2\right)=1$.
Following these definitions we define the generalized omega function
for the I\textendash LEC as,

\begin{equation}
\Omega_{IG}=E_{Iue}-E_{Ilu},\label{oig}
\end{equation}
where $E_{Iue}=\overset{.}{Q_{cI}}$, because $E_{u}=J_{I2}=\overset{.}{Q_{cI}}$
and the minimum performance of the I\textendash LEC is $r_{min}=0$,
and $E_{lu}=\epsilon_{M\epsilon}T_{h}J_{I1}X_{I1}-J_{I2}$, with the
maximum performance $r_{max}=\epsilon_{M\epsilon}$ and $E_{i}=T_{h}J_{I1}X_{I1}$.
Substituting the fluxes and forces (Eqs. \ref{ji1} and \ref{ji2})
in $E_{Iue}$ and $E_{Ilu}$ and the inverse force ratio (Eq. \ref{xi}),
we obtain the generalized omega function for the I\textendash LEC,

\begin{equation}
\Omega_{IG}^{\#}\left(x_{I},q,\epsilon_{C}\right)=-\epsilon_{C}\left[2\left(\frac{q+x_{I}}{x_{I}}\right)+\left(\frac{q}{1+\sqrt{1-q^{2}}}\right)^{2}\left(\frac{1+qx_{I}}{x_{I}^{2}}\right)\right],\label{oigx}
\end{equation}
here we used the factor of normalization $T_{h}L_{22}X_{I2}^{2}$. 

In Fig. \ref{fig3}a we observe that this function could give us the
$M\Omega_{IG}$\textendash working regime by solving $\left.\partial_{x_{I}}\Omega_{IG}^{\#}\left(x_{I},q,\epsilon_{C}\right)\right|_{x_{M\Omega_{IG}}}=0$,
and obtain the inverse force ratio for this regime, 
\begin{equation}
x_{M\varOmega_{IG}}\left(q\right)=\frac{2q}{q^{2}-4\left(1+\sqrt{1-q^{2}}\right)},\label{xmoig}
\end{equation}
to evaluate the characteristic functions and get the energetics of
the I\textendash LEC working in this regime shown in Fig. \ref{fig3}b.

\subsubsection{Energetics of the I\textendash LEC}

The energetics of the I\textendash LEC is shown in Figure \ref{fig3}.
In this case we observe $x_{ME_{IG}}=x_{M\Omega_{IG}}$as for the
D\textendash LEC case, therefore the characteristic functions in these
working regimes are the same. The functions that describe the energetics
of the I\textendash LEC are shown in Table \ref{t2}.

We can see from the Figure \ref{fig3} the comparison between the
optimization criteria ($Imdf$, $M\left(E,\Omega\right)_{IG}$, $M\epsilon$,
$\dot{Q}_{cI}^{\#}$). We note that the COP of this non-isothermal
linear converter working in the various operating regimes, keeps the
following hierarchy $\epsilon_{C}>\epsilon_{M\epsilon}>\epsilon_{M\left(E,\Omega\right)_{IG}}$,
in the same way as the same converter; we observe the hierarchy of
dissipation and cooling load (see Figures \ref{fig3}b and \ref{fig3}d)
under different operating modes: $\varPhi_{IM\left(E,\Omega\right)_{DG}}^{\#}>\varPhi_{IM\epsilon}^{\#}>\varPhi_{Imdf}^{\#}$
and $\dot{Q}_{cIM\left(E,\Omega\right)_{DG}}^{\#}>\dot{Q}_{cIM\epsilon}^{\#}$,
respectively.

\section{Thermoelectric Thomson's relations for a non\textendash isothermic
LEC\label{rel}}

\begin{figure}[t]
\centering a) %
\fbox{\includegraphics[width=15cm,height=7.5cm]{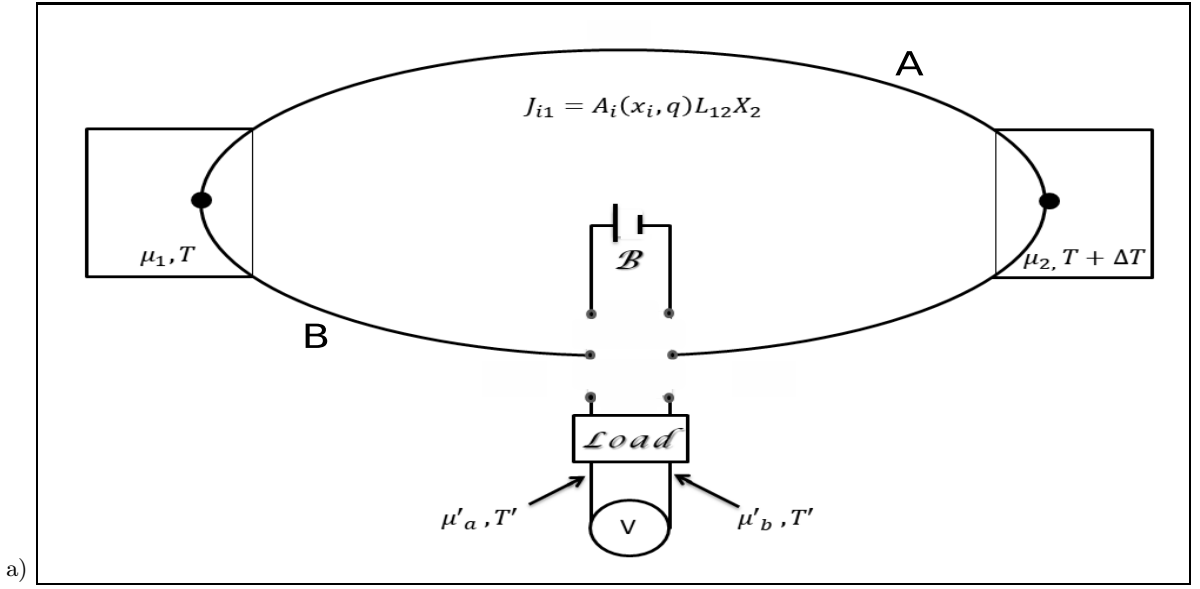} } \caption{a) Outline of a thermocouple with a variable electric current $J_{i1}=A_{i}(x_{ij},q)L_{12}\nabla\frac{1}{T}$,
where $A_{i=D,I}(x_{ij},q)=\left\{ \left(1+\frac{x_{Dj}}{q}\right),\left(1+\frac{1}{qx_{Ij}}\right)\right\} $
and $j=M\eta,\;M\left(E,\Omega\right)_{DG},\;MP_{D},\;M\left(E,\Omega\right)_{IG}$
or $M\epsilon$, respectively. These charge fluxes can be tuned\textendash in
with a non\textendash resistive load for a D\textendash LEC or a battery
for an I\textendash LEC.}
\label{fig4} 
\end{figure}

In this section we will make a proposal to introduce several working
regimes in the thermoelectric phenomena theory, constructed within
the LIT. We use our previous models of a steady linear energy converter
(D\textendash LEC \& I\textendash LEC), for small $\triangle T$,
to describe a thermocouple subject to a heat flux, given by $J_{i2}=J_{Q}$
(Eq. \ref{jd2}b for $i=D$ or Eq. \ref{ji2}b for $i=I$) and a charge
flux $J_{i1}=-J_{N}$ (Eq. \ref{jd1}b for $i=D$ or Eq. \ref{ji1}b
for $i=I$). We will introduce these working regimes through the use
of Eq.\ref{pq} as the revisited Seebeck power $\xi=\left.\partial_{T}V\right|_{J_{i1}}$.
Also, we will place a non-resistive load (the system transfer work
to the surroundings) or a battery (the surroundings transfer work
to the system) in the thermocouple at temperature $T'$, between the
points $a$ and $b$; with these elements we could tune\textendash in
the flux $J_{i1}$ for each operation mode (see Fig.\ref{fig4}).
 For example, in the case of the minimum dissipative mode the current
$J_{D1}(q,x_{mdf})=0$ (D\textendash LEC), so we will have a load
such that it allows no passage of electric current but offers no resistance
to the heat flow \cite{callen85}.

Now we will deduce the two phenomenological coefficients from the
definitions of the electric $\left(c\right)$ and heat $\left(\kappa\right)$
conductivities. The conductivity $c$ is defined as the electric flux
per unit potential gradient in an isothermal system ($\nabla T=0$).
Additionally, if the system is homogeneous then $\nabla\mu=\nabla\mu_{e}$
substituting these conditions in the generalized equations for the
thermocouple (Eqs. \ref{egtp}a), we obtain,
\begin{equation}
c=\frac{e^{2}L_{11}}{T}\Rightarrow L_{11}=\frac{cT}{e^{2}},\label{econ}
\end{equation}
where $e$ is the electric charge. The heat conductivity $\kappa$
is defined as the heat flux per unit temperature gradient for zero
electric field in an homogeneous medium, introducing this definition
in Eqs. (\ref{egtp}) we get,
\begin{equation}
\kappa=\frac{L_{22}}{T^{2}}\Rightarrow L_{22}=\kappa T^{2}.\label{tcon}
\end{equation}
These direct coefficients correspond to the well known phenomelogical
laws, Ohm's law and Fourier's law.

\subsection{Second Thomson's relation}

On the other hand we will perform a procedure to obtain the cross
coefficients. First, we consider a flux $J_{i1}=-J_{N}$ of the form
$J_{i1}\left(x_{ij},q\right)=A_{i}(x_{ij},q)L_{12}\nabla\frac{1}{T}$
(see Eqs. \ref{jd1}b and \ref{ji1}b), where,
\begin{equation}
A_{i}(x_{ij},q)=\begin{cases}
1+\frac{x_{Dj}}{q},\quad for & i=D\\
1+\frac{1}{qx_{Ij}},\quad for & i=I
\end{cases},\label{a}
\end{equation}
and $j=M\eta,\;M\left(E,\Omega\right)_{DG},\;MP_{D},\;M\left(E,\Omega\right)_{IG}$
or $M\epsilon$ respectively. Replacing $J_{i1}$ in (\ref{egtp}a)
we have
\begin{equation}
\nabla\mu=-\frac{\left[A_{i}(x_{ij},q)-1\right]L_{12}}{TL_{11}}\nabla T\label{pots}
\end{equation}

The Seebeck effect is the phenomenon that consists of the production
of an electromotive force emf in a thermocouple under the condition
of a null electric current $J_{N}=0$. For our proposal to introduce
different modes of operation we will consider a flux $J_{i1}$ compatible
with each mode.

Now integrating and rewriting Eq. (\ref{pots}) in terms of $\mu_{a}^{\text{\textasciiacute}}$
and $\mu_{b}^{\text{\textasciiacute}}$ see (Fig. \ref{fig4}) we
obtain

\begin{equation}
\mu_{b-}^{\text{\textasciiacute}}\mu_{a}^{\text{\textasciiacute}}=-\left[A_{i}(x_{ij},q)-1\right]\intop_{1}^{2}\left(\frac{L_{12}^{A}}{TL_{11}^{A}}-\frac{L_{12}^{B}}{TL_{11}^{B}}\right)dT,\label{pq}
\end{equation}
rewrite \ref{pq} in terms of $\mu_{1}$ and $\mu_{2}$ we obtain
the following,

\begin{equation}
\mu_{b-}^{\text{\textasciiacute}}\mu_{a}^{\text{\textasciiacute}}=-\left[A_{i}(x_{ij},q)-1\right]\left(\mu_{2-}\mu_{1}\right).
\end{equation}
But, because there is no temperature gradient across the voltmeter,
the voltage is given as follows

\begin{equation}
V=\frac{1}{e}\left(\mu_{b-}^{\text{\textasciiacute}}\mu_{a}^{\text{\textasciiacute}}\right)=-\frac{1}{e}\left[A_{i}(x_{ij},q)-1\right]\left(\mu_{2-}\mu_{1}\right)=-\left[A_{i}(x_{ij},q)-1\right]\intop_{1}^{2}\left(\frac{L_{12}^{A}}{eTL_{11}^{A}}-\frac{L_{12}^{B}}{eTL_{11}^{B}}\right)dT,
\end{equation}
the thermoelectric power (Seebeck power) under the condition $J_{i1}\left(x_{mdf},q\right)=0$
is defined as follows

\begin{equation}
\xi_{AB}^{mdf}=\left.\frac{\partial V}{\partial T}\right|_{J_{i1}=0}=-\left[\left(-\frac{L_{12}^{B}}{eTL_{11}^{B}}\right)-\left(-\frac{L_{12}^{A}}{eTL_{11}^{A}}\right)\right],\label{eq:37}
\end{equation}
in the same way we define the new Seebeck power, for the general case
$J_{i1}=J_{i1}\left(x_{ij},q\right)$:

\begin{equation}
\xi_{AB}=\left.\frac{\partial V}{\partial T}\right|_{J_{i1}}=-\left[A_{i}(x_{ij},q)-1\right]\left[\left(-\frac{L_{12}^{B}}{eTL_{11}^{B}}\right)-\left(-\frac{L_{12}^{A}}{eTL_{11}^{A}}\right)\right].\label{eq:38}
\end{equation}
The absolute Seebeck power is defined as

\begin{equation}
\xi\equiv-\left[A_{i}(x_{ij},q)-1\right]\frac{L_{12}^{A}}{eTL_{11}^{A}}.\label{potsm}
\end{equation}

Now we have been able to calculate the values of the phenomenological
coefficients, which have remained in terms of the Seebeck power (\ref{potsm}),
the electric conductivity (\ref{econ}), the thermal conductivity
(\ref{tcon}) and the operating constant $A_{i}$ (Eq. \ref{a}),

\begin{equation}
L_{11}=\frac{Tc}{e^{2}},\:L_{12}=-\frac{T^{2}c\xi}{\left[A_{i}(x_{ij},q)-1\right]e},\:L_{22}=T^{2}\kappa.
\end{equation}

If we accept the electrical conductivity $c$, the thermal conductivity
$\kappa$ and the absolute thermoelectric power $\xi$ as the three
physically significant dynamic properties of a medium in addition
to force ratio and the coupling parameter, we can eliminate the three
phenomenological coefficients and therefore rewrite the kinetic equations
(\ref{egtp}) as follows:

\begin{equation}
-J_{N}=\left(\frac{Tc}{e^{2}}\right)\frac{1}{T}\nabla\mu+\left\{ -\frac{T^{2}c\xi}{\left[A_{i}(x_{ij},q)-1\right]e}\right\} \nabla\frac{1}{T},\label{eq:3-1-1}
\end{equation}

\begin{equation}
J_{Q}=\left\{ -\frac{T^{2}c\xi}{\left[A_{i}(x_{ij},q)-1\right]e}\right\} \frac{1}{T}\nabla\mu+T^{2}\kappa\nabla\frac{1}{T},\label{eq:4-1}
\end{equation}

As is well known, the Peltier effect describes the way in which the
heat of an isothermal welding $\left(\nabla T=0\right)$ produced
by an electric current evolves. Under the condition that the process
of heat evolution in the welding is isothermal, the dynamic equations
(\ref{egtp}) take the following form,

\begin{equation}
J_{Q}=-\frac{T\xi}{\left[A_{i}(x_{i},q)-1\right]}\left(eJ_{N}\right),
\end{equation}
where $J_{Q}=J_{Q}^{B}-J_{Q}^{A}=-\frac{T\left(\xi{}_{B}-\xi_{A}\right)}{\left[A_{i}(x_{i},q)-1\right]}\left(eJ_{N}\right)$.
On the other hand the Peltier coefficient $\pi_{AB}$ is defined as

\begin{equation}
\pi_{AB}=\frac{J_{Q}}{eJ_{N}}=-\frac{T\left(\xi{}_{B}-\xi_{A}\right)}{\left[A_{i}(x_{i},q)-1\right]}=-\frac{T\xi_{AB}}{\left[A_{i}(x_{i},q)-1\right]},\label{eq:45}
\end{equation}
this last relation is called the second Thomson's relation, which
shows a subtle relation between the Seebeck power and the Peltier
coefficient.

\subsection{First Thomson's relation}

In this section we will make a deduction of the first Thomson's relation,
for which we will proceed in a habitual way \cite{callen85,garcia03}.
We will begin considering the Fig. \ref{fig1}a which is a synthesized
description of a thermocouple. We must fix our attention in the soldering;
suppose that there is a charge unit transfer along the thermocouple
which inevitably causes several transfers of energy in the system,
the analysis of these energy transfers will help us to construct an
energy balance equation from which we obtain the first Thomson's relation.

Before starting with the analysis of the energy balance it is necessary
to define the Thomson coefficient $\tau$ which is defined as the
Thomson heat absorbed per unit temperature gradient and per unit electric
current

\begin{equation}
\tau\equiv\frac{Thomson\:heat}{\left(eJ_{N}\right)\nabla T}=T\frac{d\xi}{dT}
\end{equation}

Consider that the load unit passes the welding to temperature $T$
in the clockwise direction from $B$ to $A$; this causes a heat to
be absorbed from the source due to the Peltier effect $\pi_{BA}$,
the load now on the material $A$ absorbs a heat of Thomson $\tau_{A}dT$,
the load follows its path through the circuit in such a way that,
at the time of traversing the welding that lies at ($T+dT$) in a
clockwise manner from $A$ to $B$ in such a way that the system absorbs
Peltier heat $(\pi_{AB}+d\pi_{AB})$ the charge in its path traverses
material $B$ where it absorbs a heat due to the Thomson effect $(dV=-dV)$.
Finally, when the charge crosses the battery performs a work equal
to the $emf$ that produces the battery ($dV=-dV$). We must mention
that we have not considered the contributions of heat in the balance
due to the heat of Joule since in our analysis this is small in comparison
to contributions due to the heat of Thomson.

Now if we equalize the total energy that is absorbed by the system
along the path of the circuit with the work done on the battery we
get the following

\begin{equation}
-\pi_{AB}+\left(\pi_{AB}+d\pi_{AB}\right)+\left(\tau_{A}-\tau_{B}\right)dT=dV,
\end{equation}
which can be rewritten as follows

\begin{equation}
\tau_{A}-\tau_{B}=-\frac{d\pi}{dT}+\frac{dV}{dT}
\end{equation}
using (\ref{eq:37}) and (\ref{eq:45})

\begin{equation}
\frac{dV}{dT}=\xi_{AB}=-\left[A_{i}(x_{i},q)-1\right]\frac{\pi_{BA}}{T},\label{eq:48}
\end{equation}
therefore

\begin{equation}
\frac{d\pi_{AB}}{dT}+\left(\tau_{A}-\tau_{B}\right)=\xi_{AB},
\end{equation}
or

\begin{equation}
\frac{d\pi_{AB}}{dT}+\left(\tau_{A}-\tau_{B}\right)=-\left[A_{i}(x_{i},q)-1\right]\frac{\pi_{AB}}{T},
\end{equation}

Which is the first Thomson's relation for any mode of operation, when
$x_{mfd}=-q$ (minimum dissipation function) reproduces the first
Thomson's relation (see Table .\ref{tab3}).

{\tiny{}}
\begin{table}[tb]
{\tiny{}\centering}{\scriptsize{}}%
\begin{tabular}{clcccc}
\toprule 
\multirow{2}{*}{Relation} & D\textendash LEC &  &  &  & \tabularnewline
 & working regimes & $x_{mdf}$ & $x_{M\eta}$ & $x_{MP_{D}}$ & $x_{M\left(E,\Omega\right)_{DG}}$\tabularnewline
\midrule 
FTR &  $\frac{d\pi_{AB}}{dT}+\left(\tau_{A}-\tau_{B}\right)=$ & $\frac{\pi_{AB}}{T}$ & $\frac{1}{1+\sqrt{1-q^{2}}}\frac{\pi_{AB}}{T}$ & $\frac{1}{2}\frac{\pi_{AB}}{T}$ & $\frac{q\left(q^{2}-4\right)}{4\left(q^{2}-2\right)}\frac{\pi_{AB}}{T}$\tabularnewline
\midrule 
STR  & $\pi_{AB}=$ & $T\xi_{AB}$ & $\left(1+\sqrt{1-q^{2}}\right)T\xi_{AB}$ & $2T\xi_{AB}$ & $\frac{4\left(q^{2}-2\right)}{q\left(q^{2}-4\right)}T\xi_{AB}$\tabularnewline
\midrule 
\multirow{2}{*}{} & I\textendash LEC &  &  &  & \tabularnewline
 & working regimes & $x_{I_{mdf}}$ & $x_{M\epsilon}=x_{M\eta}$ & $x_{M\left(E,\Omega\right)_{IG}}$ & \tabularnewline
\midrule 
FTR &  $\frac{d\pi_{AB}}{dT}+\left(\tau_{A}-\tau_{B}\right)=$ & $\frac{1}{q^{2}}\frac{\pi_{AB}}{T}$ & $\frac{1}{1+\sqrt{1-q^{2}}}\frac{\pi_{AB}}{T}$ & $-\left[\frac{2}{q^{2}-4\left(1+\sqrt{1-q^{2}}\right)}\right]\frac{\pi_{AB}}{T}$ & \tabularnewline
\midrule 
STR &  $\pi_{AB}=$ & $q^{2}T\xi_{AB}$ & $\left(1+\sqrt{1-q^{2}}\right)T\xi_{AB}$ & $-\left[\frac{q^{2}-4\left(1+\sqrt{1-q^{2}}\right)}{2}\right]T\xi_{AB}$ & \tabularnewline
\bottomrule
\end{tabular}{\scriptsize{} }{\tiny{}\caption{The table shows the first and the second Thomson's relations (FTR)
and (STR) evaluated in optimal force ratio of different objective
functions, in the case when the system operates as a D\textendash LEC,
the evaluation points are minimum dissipation function $x_{mdf}$,
maximum efficiency $x_{M\eta}$, maximum power output $x_{MP_{D}}$,
and the point of maximum ecological function which is equivalent to
the maximum omega function $x_{M(E,\varOmega)_{DG}}$. Similary when
the system operates as a I\textendash LEC, the force ratio from which
the Thomson's relations will be evaluated are the minimum dissipation
function $x_{I_{mdf}}$, maximum COP $x_{M\epsilon}$ and the force
ratio of maximum ecological function and maximum omega function $x_{M(E,\varOmega)_{IG}}$.}
}{\tiny \par}

{\tiny{}\label{tab3} }{\tiny \par}
\end{table}
{\tiny \par}

\section{Concluding remarks\label{conc}}

For the deduction of two Thomson's relations, usually two experiments
are carried out which imply that the thermocouple transfers energy
in the form of work to its surroundings (Seebeck effect) or receives
energy in the form of work of these (Peltier effect). In order to
build a model in the context of linear irreversible thermodynamics,
which takes into account these exchanges of work, here we proposed
a non\textendash isothermal energy converter, since of the fluxs involved
in this phenomenon one of them is of heat and occurs between two heat
reservoirs whose temperatures are fixed ($T$ and $T+\triangle T$).
This converter works in two modes, such as a heat engine where a spontaneous
heat flux promotes a non\textendash spontaneous (D-LEC), and as a
refrigerator where a spontaneous flow of any nature promotes a non\textendash spontaneous
heat flow (I -LEC). For these two converters we find different working
regimes (steady states) that correspond to a specific relation between
the force ratios (operation) and the coupling coefficient (design).

This fact allows that for each of given regimes, an expression for
$J_{D1}$ and the force $X_{D2}$ (Eq. \ref{jd1}b) or, where appropriate,
for $J_{I1}$ and the force $X_{I2}$ (Eq. \ref{ji1}b), so as to
ensure that reciprocity relationships are satisfied in each regime.
Thus, we propose that the extra-thermodynamic information required
to find the cross-coefficients in any system, given the linear relation
between fluxs and forces and the bilinear form of entropy production
\cite{Yourgrau}, can be obtained from the relation between the force
ratio $x_{i}$ and the coupling coefficient $q$ that provides the
thermodynamic optimization. From the above we can say that the direct
coefficients can be deduced from microscopic models or from phenomenological
laws such as Ohm's law and Fourier's law, while the cross\textendash coefficients
necessarily come from an experiment that involves the interaction
of the system with the surroundings .

When applying these results to the thermoelectric phenomena, we find
that for the working regime that corresponds to a minimum dissipation
function of the thermocouple, operating as D-LEC, recovers the already
known second Thomson's relation (see Table \ref{tab3}). This is so
because the dissipation in this regime is minimal and therefore the
production of entropy also, in fact can be verified that in this regime
$J_{D1}\left(x_{mdf},q\right)=0$, that it is precisely the condition
of open circuit that is used to deduce this relation. Similarly, for
the regime of minimum dissipation function, we recover the well known
first Thomson's relation. Additionally, a new set of Thomson relations
is obtained; this comes from the thermodynamic optimization, which
provides information on how to transfer the system work to the surroundings,
for example, in the case of the work regime of maximum output power
of the D-LEC, the second relation given by $\pi_{AB}=2T\xi_{AB}$
is obtained, while for the first relation we obtain $d\pi_{AB}/dT+\left(\tau_{A}-\tau_{B}\right)=\pi_{AB}/2T$.
These two relations are like this because a non\textendash resistive
load has been placed between points $a$ and $b$ which the thermocouple
transfers a maximum amount of work consistent with the flux of electric
charge given by $J_{D1}\left(x_{MP_{D}},q\right)=-\left(Tc/e\right)\left(\nabla\mu/eT\right)$
(see Eq. \ref{ji1}a). In the case of the Peltier effect (I-LEC) at
maximum generalized ecological function, the current that must force
the battery (external work) in the thermocouple must be, $J_{I1}\left(x_{MP_{D}},q\right)=\left(1/3\right)\left(Tc/e\right)\left(\nabla\mu/eT\right)$
(see Eq. \ref{ji1}), at the strong coupling condition. Applying systematically
the energetics of the linear stationary energy converter developed
in Section \ref{lec}, we obtain 6 new Thomson's \textquotedbl{}second
relations\textquotedbl{} and as many Thomson's \textquotedbl{}first
relations\textquotedbl{}. In fact, from the results shown in Table
\ref{tab3}, we see that in the condition of strong coupling, the
Peltier heat in the different working regimes satisfies $\pi_{AB}^{mdf}=\pi_{AB}^{I_{mdf}}=\pi_{AB}^{M\eta}=\pi_{AB}^{M\epsilon}<\pi_{AB}^{M\left(E,\Omega\right)_{DG}}<\pi_{AB}^{M\left(E,\Omega\right)_{IG}}<\pi_{AB}^{MP_{D}}$.

Then the concept of energy converter, isothermal \cite{valencia17}
and non-isothermal, developed from the division of the production
of entropy into two subsets of products, one made up of the products
of fluxes and forces that contribute positively to this, and the other
formed by those products that contribute to it negatively, it can
be useful to explore the cross-contributions to the flows that intervene
in the system (Eqs. \ref{jd1}, \ref{jd2}, \ref{ji1} and \ref{ji2})
as has been shown in this article. Another example is found in the
reference mentioned above, where it is shown that given the elements
of an electrical circuit and the different work regimes there is a
specific relationship between them. This allows us to affirm that
the use of optimization criteria developed for other converter models
\cite{feng} in other contexts opens up the possibility of designing
new experiments within linear irreversible thermodynamics.

\subsection*{Acknowledgement}

We thank RMMA and FAB for stimulating discussions, suggestions and
invaluable help in the preparation of the manuscript. This work was
supported in part by EDI\textendash SIP\textendash COFAA\textendash IPN,
CONACYT, Mexico.

\section*{Appendix}

The efficient power $P_{\eta}$ is defined as $P_{\eta}=P\eta$, which
can be written in terms of generalized fluxs and forces as follows,

\begin{equation}
P_{\eta}=\frac{\left(T_{c}J_{D1}X_{D1}\right)^{2}}{J_{D2}}=\eta_{c}^{2}\frac{\left[x\left(x+q\right)\right]^{2}}{\left(qx+1\right)}L_{22}X_{2},\label{eq:13-1}
\end{equation}
or

\begin{equation}
P_{\eta}^{*}=\eta_{c}\frac{\left[x\left(x+q\right)\right]^{2}}{\left(qx+1\right)}.
\end{equation}

\begin{figure}
\begin{centering}
\includegraphics[width=7cm,height=5cm]{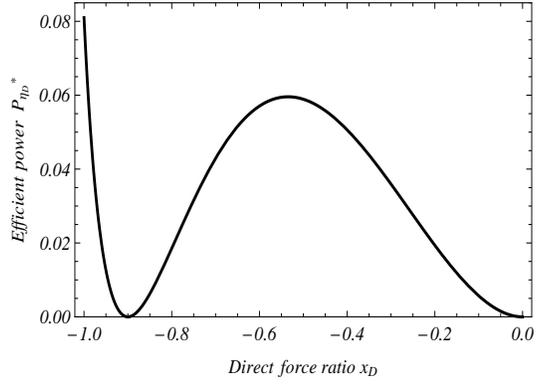}
\par\end{centering}
\caption{The figure shows the normalized Efficient Power $P_{\eta}^{*}$ for
the steady state non\textendash isothermic D\textendash LEC; here
we take $q=0.9$ and $\eta_{C}=0.9$.}

\label{potefi}
\end{figure}

Optimizing the Equation \ref{eq:13-1} with respect to $x$, it is
possible to obtain the maximun efficient power force ratio $x_{MP_{\eta}}$
(see Figure \ref{potefi}),

\begin{equation}
x_{MP_{\eta}}\left(q\right)=\frac{4}{6}\frac{\sqrt{1-q^{2}+\left(\frac{q}{2}\right)^{4}}-\left(1+\left(\frac{q}{2}\right)^{2}\right)}{q},\label{xpefi}
\end{equation}
and we derive the energetics of the D\textendash LEC from Eq. (\ref{xpefi}).

\end{document}